# Clustering genes of common evolutionary history

Kevin Gori[1], Tomasz Suchan[2], Nadir Alvarez[2], Nick Goldman[1,*] and Christophe Dessimoz[1,2,3,4,5,*]

[1]European Molecular Biology Laboratory, European Bioinformatics Institute,
Wellcome Trust Campus, Hinxton, CB10 1SD, UK
[2]Department of Ecology and Evolution, Biophore Building, UNIL-Sorge, University of Lausanne,
1015 Lausanne, Switzerland
[3]Department of Genetics, Evolution & Environment and Department of Computer Science,
University College London, Gower St, London, WC1E 6BT, UK
[4]Centre for Integrative Genomics, University of Lausanne, 1015 Lausanne, Switzerland
[5]Swiss Institute of Bioinformatics, Biophore, 1015 Lausanne, Switzerland

*Correspondence : goldman@ebi.ac.uk; c.dessimoz@ucl.ac.uk

## Abstract

Phylogenetic inference can potentially result in a more accurate tree using data from multiple loci. However, if the loci are incongruent—due to events such as incomplete lineage sorting or horizontal gene transfer—it can be misleading to infer a single tree. To address this, many previous contributions have taken a mechanistic approach, by modelling specific processes. Alternatively, one can cluster loci without assuming how these incongruencies might arise. Such "process-agnostic" approaches typically infer a tree for each locus and cluster these. There are, however, many possible combinations of tree distance and clustering methods; their comparative performance in the context of tree incongruence is largely unknown. Furthermore, because standard model selection criteria such as AIC cannot be applied to problems with a variable number of topologies, the issue of inferring the optimal number of clusters is poorly understood. Here, we perform a large-scale simulation study of phylogenetic distances and clustering methods to infer loci of common evolutionary history. We observe that the best-performing combinations are distances accounting for branch lengths followed by spectral clustering or Ward's method. We also introduce two statistical tests to infer the optimal number of clusters and show that they strongly outperform the silhouette criterion, a general-purpose heuristic. We illustrate the usefulness of the approach by (*i*) identifying errors in a previous phylogenetic analysis of yeast species and (*ii*) identifying topological incongruence among newly sequenced loci of the globeflower fly genus *Chiastocheta*. We release treeCl, a new program to cluster genes of common evolutionary history (http://git.io/treeCl).

# Introduction

Molecular phylogenetic methods infer the evolutionary history of homologous sequences. The techniques of molecular phylogenetics were developed in the analysis of individual protein sequences (Neyman 1971; Kashyap and Subas 1974), but due to the modern abundance of sequencing data it is increasingly common to infer trees by jointly analysing sequences from multiple loci (Delsuc et al. 2005). By considering more data, multilocus analyses are expected to deliver better-resolved and less biased inferences by averaging out uncertainty over a greater amount of data (Pamilo and Nei 1988).

There are a number of methods for multilocus phylogenetic analysis (Bininda-Emonds et al. 2002; de Queiroz and Gatesy 2007; Liu et al. 2009). Many of these proceed by inferring the single evolutionary tree that best fits the entire data set. Such "averaging" over multiple loci presumes that these loci share a common evolutionary history. However, when a data set comprises multiple loci, the trees derived from individual loci have the potential to be incongruent (Jeffroy et al. 2006). A key question here is whether incongruence results from sampling error, or if it indicates a real underlying difference in the evolution of distinct genomic loci. If we build a single summary tree from multiple loci we are implicitly assuming the former: that each locus is a noisy estimate of the same underlying tree.

Alternatively, we might expect different regions of a genome to have different histories (Leigh, Lapointe, et al. 2011), due to a variety of processes such as horizontal gene transfer (HGT), hybridisation, incomplete lineage sorting (ILS) and recombination. If we believe such processes have occurred, then we should expect that the trees derived from different loci could be incongruent with one another. Consequently, "summary" trees inferred from the entire data set may be only partially representative or, in the worst case, not representative of the evolution of any locus. Because this is a systematic error, rather than noise, we cannot expect it to be reduced by adding more data (Philippe et al. 2011). If we believe there is real heterogeneity in the evolutionary process that produced the genomes, and incongruence is an indication of this, then we should look for ways of partitioning multilocus data into groups that are related by the same history (Bull et al. 1993; Huelsenbeck et al. 1994; Cunningham 1997; Waddell et al. 2000).

Many methods dealing with incongruence make explicit assumptions about its biological basis.

Such "mechanistic" approaches have been developed to model HGT (Hallett and Lagergren 2001; Dessimoz et al. 2008; Abby et al. 2010), ILS (Rannala and Yang 2003; Heled and Drummond 2010), recombination (Kosakovsky Pond et al. 2006), gene duplication (GD) (Chen et al. 2000; Boussau et al. 2013), and combinations of processes such as combined ILS/GD models (Bansal et al. 2010; Doyon et al. 2010; Szöllősi and Daubin 2012). However, mechanistic approaches can be computationally prohibitive, and may not be robust to other unmodelled sources of incongruence.

We focus our attention on an alternative class of methods that we will describe as "process-agnostic". These aim to detect the existence and extent of any significant incongruence within a data set, without relying on any assumptions about its biological basis. Existing process-agnostic approaches take the form of statistical tests of incongruence (Planet 2006; Leigh, Lapointe, et al. 2011) and clustering approaches relying on partitioning data sets into groups that are cohesive and self-similar (Nye 2008; Leigh, Schliep, et al. 2011).

Nye's Tree of Trees (2008) summarises the phylogenetic similarities among genes as another tree, termed a meta-tree, where a tip corresponds to a tree derived from multilocus data, and an internal node represents the consensus of its child trees. The meta-tree is inferred from inter-tree Robinson-Foulds (1981) distances using an algorithm analogous to neighbour-joining (Saitou and Nei 1987).

Similarly, Conclustador (Leigh, Schliep, et al. 2011) uses inter-tree distances as a basis for clustering. Trees are compared using a novel Euclidean distance among bipartitions weighted by bootstrap support, and for clustering Leigh et al. use a version of the $k$-means algorithm and a spectral clustering method (Kaufman and Rousseeuw 1987; Zelnik-Manor and Perona 2004). A conceptually similar method is PhyBin (Newton and Newton 2013), which can either identify genes with topologically identical trees or perform hierarchical clustering on the Robinson-Foulds distance matrix between every tree.

Statistical binning (Mirarab et al. 2014) uses a graph-based algorithm to divide a set of genes into a number of approximately equal-sized bins of phylogenetically compatible genes (Warnow 1994). This has been used as a preprocessing step, with the bins subsequently used as input for coalescent species tree estimation; binning is shown to reduce run times, and to increase accuracy in the presence of ILS (Mirarab et al. 2014).

BUCKy (Ané et al. 2007; Larget et al. 2010) uses a Bayesian probabilistic framework to estimate a gene-to-tree map that assigns each gene to one of the $(2n - 3)!!$ possible unrooted trees on $n$ taxa (Felsenstein 2004). A Dirichlet process prior (Ferguson 1973; Antoniak 1974) is used to determine the total number of distinct trees represented by the gene-to-tree map.

These methods have in common that they each adopt a specific clustering procedure. There are, however, many potential distance measures and clustering algorithms, and we know almost nothing about their relative performance in identifying genes that share common evolutionary histories under plausible biological scenarios. For instance, the Robinson-Foulds distance used in Tree of Trees ignores any difference in branch lengths among trees, yet these might provide useful information in the context of ILS; the Dirichlet process prior in BUCKy tends to result in uneven cluster sizes (Ané et al. 2007), yet this might be suboptimal in the context of recombination. Furthermore, the problem of determining the optimal number of clusters remains poorly understood, with methods providing no, or only generic, solutions.

Here, we present a survey of clustering methods to partition multilocus data sets into groups with consistent underlying phylogenies. Our aims are to investigate whether this is a viable approach to use to partition multilocus data in an evolutionarily meaningful way, and to measure the relative effectiveness of each method. Specifically, we test combinations of three distance measures between trees (table 1) and seven well-established clustering algorithms (table 2) on simulated and empirical sequence data.

We also introduce two likelihood ratio tests for inferring the optimal number of clusters. We test them extensively through simulations and show that they accurately recover the true number of clusters and outperform the silhouette criterion, a general-purpose heuristic.

We apply the best combination of tree distance, clustering method and stopping criterion to two empirical data sets: alignments of 344 loci in 18 yeast taxa (Hess and Goldman 2011), and of 176 loci of 306 taxa derived from 7 species of *Chiastocheta* genus globeflower flies.

The analyses were carried out using our new open source software package, treeCl, freely available at http://git.io/treeCl.

# Results

The clustering approach investigated here takes a set of sequence alignments (one alignment per locus), and from them describes a partition of the data that divides the alignments into non-overlapping subsets, each subset containing loci sharing a common phylogenetic history. Throughout this paper we will describe such a division as a *partition*, and the resulting subsets as *clusters*. The approach is a three-step pipeline (Figure 1). First, we infer a separate phylogenetic tree for each input sequence alignment. Second, we gauge the level of evolutionary similarity among loci by measuring distances between pairs of trees. Third, we apply a clustering algorithm on the distances to generate a set of clusters. The number of clusters is either a fixed value decided *a priori*, or inferred from the data using tests introduced below.

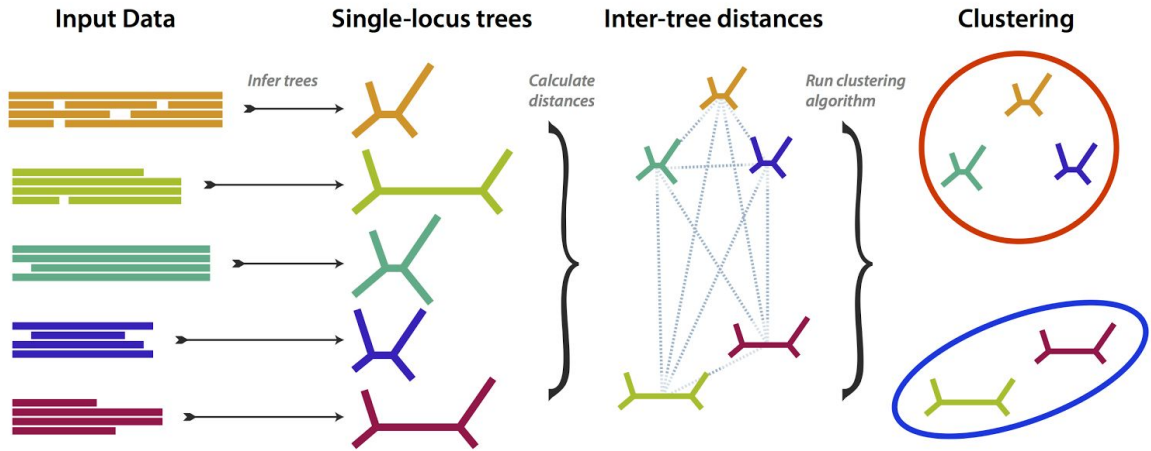

*Figure 1: Overview of the clustering process. From left to right: input alignments are read; trees are inferred from the alignments; inter-tree distances are computed and used as the basis for clustering. Further procedures are used to re-estimate one tree for each cluster and to choose the optimal number of clusters—see text for details.*

In the following, we describe the results of a series of simulation experiments designed to explore the parameter space of the tree clustering approach and choose the most effective combinations of methods. We assess different stopping criteria for choosing the best-supported number of clusters from the data, again using simulation. Finally, we present the application of our method to data sets of yeast orthologs and of *Chiastocheta* genus globeflower flies.

## Performance of the combinations of distance metrics and clustering methods.

Combinations of clustering methods (table 1) and distance metrics (table 2) were tested on simulated data over a range of conditions, described in *Materials and Methods* (table 3).

We investigated the performance of combinations of distance metrics and clustering methods for a fixed and known number of clusters. To assess the accuracy of each resulting partition, we computed the difference between the true partition (known from simulation) and the inferred partition using variation of information, an information-theoretic measure of the difference between two partitions of the same set (Meilă 2007). A variation of information value of zero is obtained when the two partitions are the same, and increasing positive values are obtained for partitions that are increasingly different.

Our results are summarised in Figure 2. In terms of distance metrics, the performance using the Euclidean and geodesic distances is considerably better than Robinson-Foulds. Of these two, the geodesic distance performs marginally better than Euclidean. These conclusions hold for both skewed and uniform cluster size distributions, for the small and large data sets (Supplementary Figures 1–3), and for scenarios simulating both ILS (using nearest-neighbour interchange rearrangements) and HGT (using subtree prune-and-regraft).

In terms of clustering methods, the performance is worst using the simpler hierarchical methods—single-linkage, complete-linkage and average-linkage. Hierarchical clustering using Ward's criterion is more successful, but the best-performing methods are those involving embedding the distance matrix in a coordinate space: spectral and multidimensional scaling (MDS). However, MDS, as well as k-medoids, shows erratic behaviour in some of the scenarios tested (Supplementary Figures 1–3), and these were not considered for further analyses. Summarising these observations, the combination of Euclidean or geodesic distances with spectral or Ward clustering seem to provide consistently the best overall performance across various conditions tested here. These combinations were used in our further analyses.

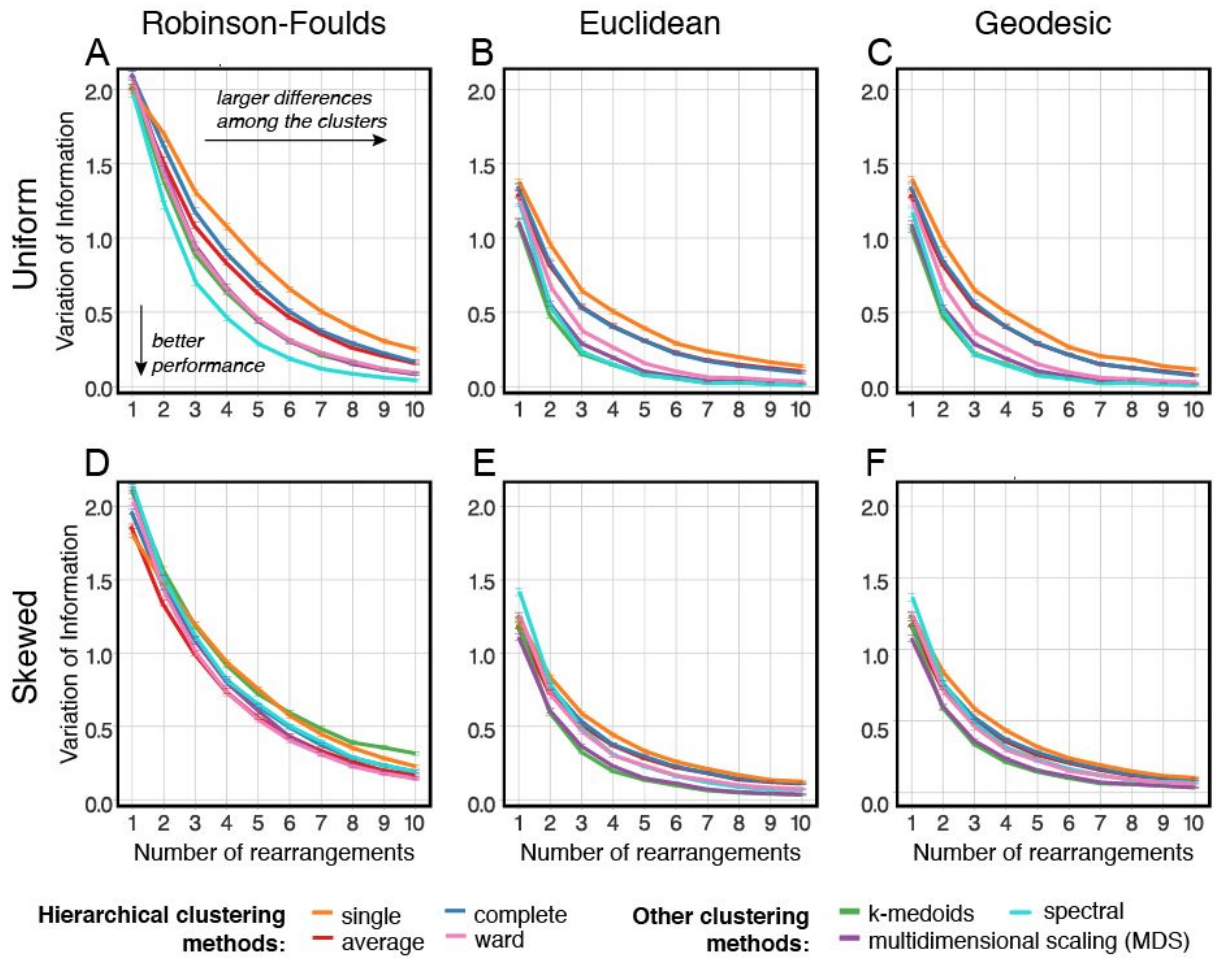

*Figure 2: The relative performances of combinations of distance metric (varying over columns of panels) and clustering methods (shown by the colours of the lines), as measured by the variation of information metric (y-axes; higher values show a larger departure from the correct solution). Lines show the mean value obtained from 1000 replicates, and the error bars show the standard error of the mean. Rows correspond to the experiments with a partition of uniformly-sized clusters (A–C) and those with a partition of clusters of skewed sizes (D–F). In each individual panel, the x-axis represents the number of NNI rearrangements separating the underlying clusters, so that increasing values along this axis correlate with the clustering problem becoming easier.*

## Performance of methods for determining the number of clusters

So far we have investigated performance with a known number of clusters, but this is typically unknown. To infer it, we devised two special-purpose likelihood ratio test procedures using empirical distributions of the test statistic: one a distribution derived from the input data via permutation, and the other derived via a parametric bootstrap resampling procedure (see *Materials and Methods*). We also compared these to a general-purpose "silhouette" criterion (Rousseeuw 1987). For a single point the silhouette value is the ratio of the mean of the distances to all other points in its cluster to the mean of the distances to all points in the nearest cluster. The silhouette score for the entire partition is the mean of these ratios over all points in the data set. The optimal number of clusters is inferred as the value for which the silhouette score is maximised.

For clarity, we first describe the results for a single set of sequences (one problem instance) before presenting our aggregate results. Given a problem instance, we repeat the clustering procedure with a varying number of clusters and compute the overall partition likelihood for each. Because specifying a greater number of clusters provides more freedom for the model to fit the data, the likelihood is expected to increase: this is generally what we observe. However, as in all likelihood ratio tests, the key consideration is by how much the likelihood must increase to warrant using the more complex model. To tackle this we generate empirical distributions of the likelihood increase from pseudo-replicate data derived from the data present in the instance, through the permutation and parametric bootstrap procedures described in *Materials and Methods*. The likelihood increase from the original data is compared to the expected increase from the empirical distribution to determine significance (Figure 3; Supplementary Figure 4).

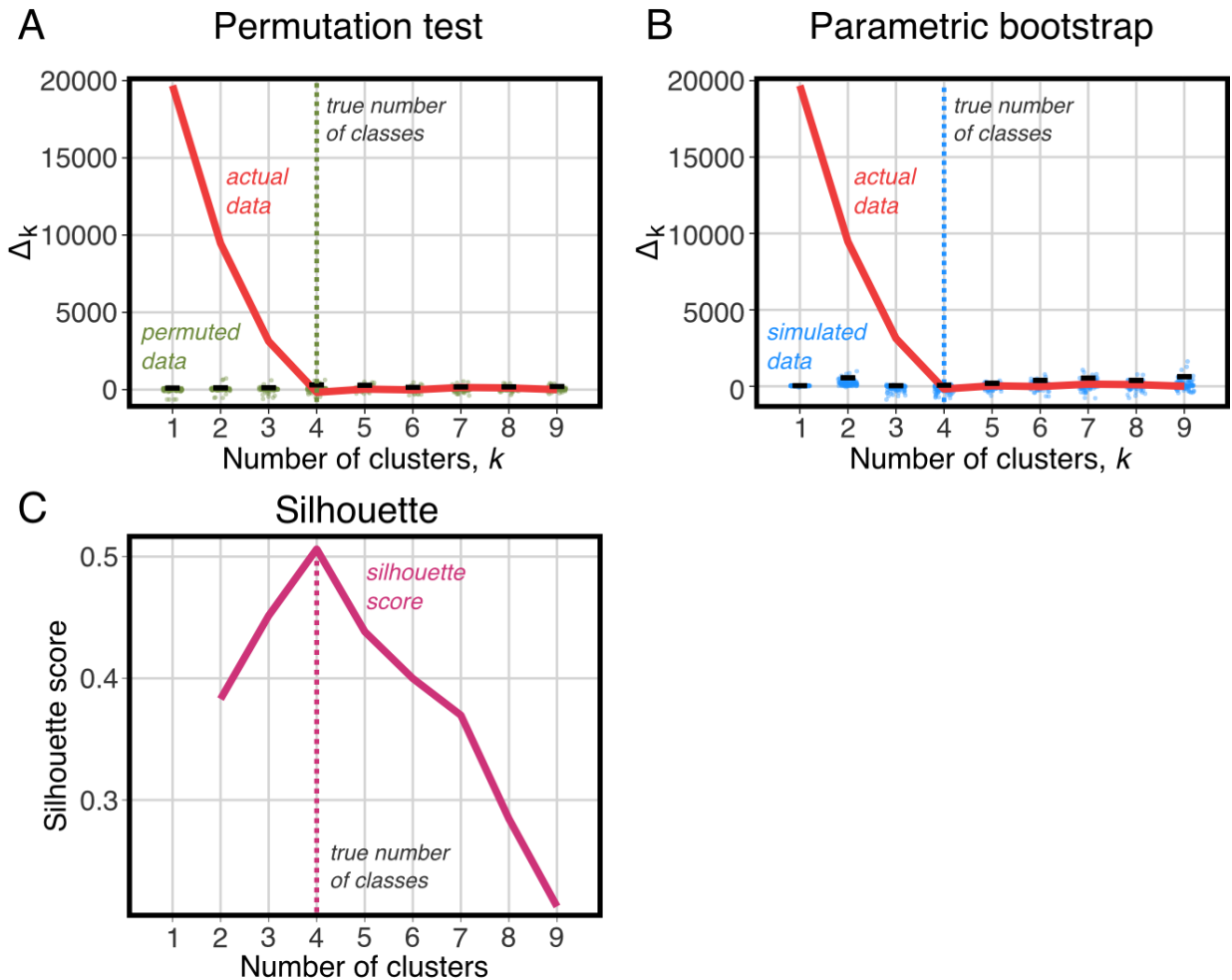

*Figure 3: Comparison of the criteria used to determine the number of clusters on a single problem instance—in this example, data simulated for 60 loci belonging to 4 clusters, each of size 15, with the clusters' trees separated by 1 SPR. As the proposed number of clusters increases, the likelihood increases, which is expected because of the greater number of free*

*parameters in the model. (A) Permutation test: the improvement in likelihood for each additional cluster (red curve) is significantly greater than that observed for permuted data sets (green dots show the distribution of values over 100 permutations) until the comparison between 4 and 5 clusters is reached, correctly implying that the use of 4 clusters is optimal. (B) Parametric bootstrap test: again, the improvement for each additional cluster (red curve) is significantly greater than that for data sets simulated for one fewer cluster (blue dots) until the true number of clusters (4) has been reached. (C) Silhouette score: the general-purpose silhouette stopping criterion has its maximum at the true value of 4. We note that in this instance, comprising a single data set from one simulation design, the three methods agree on the true answer.*

Let us now consider the results over multiple problem instances. We simulated data sets using the procedure corresponding to the "small uniform" setup (see *Materials and Methods*, sub-section "Simulating data sets with incongruence" for details), with two levels of difficulty: we generated 100 data sets from trees separated by 1 SPR move (referred to as "difficult"), and 100 separated by 5 SPR moves ("moderate"). Each data set was analysed under the four combinations of Euclidean or geodesic distances with spectral or Ward's method clustering. This resulted in a total of 800 problem instances.

To investigate the overall performance of the three stopping criteria, we first consider the aggregate results for all 400 "difficult" and 400 "moderate" problem instances, i.e. 100 each under all four combinations of distance metric and clustering procedure (Figure 4). For both the "difficult" and "moderate" cases the distribution of the number of clusters chosen is centred on the true value, 4, for all three criteria. However, in the "difficult" case, the distributions of the permutation and bootstrap tests are much tighter than the silhouette score, indicating that these two stopping criteria make correct calls more often. The results are consistent in the "moderate" case, although the differences between criteria are smaller, with all of them making many more correct calls (see also Supplementary Figure 5, E–H).

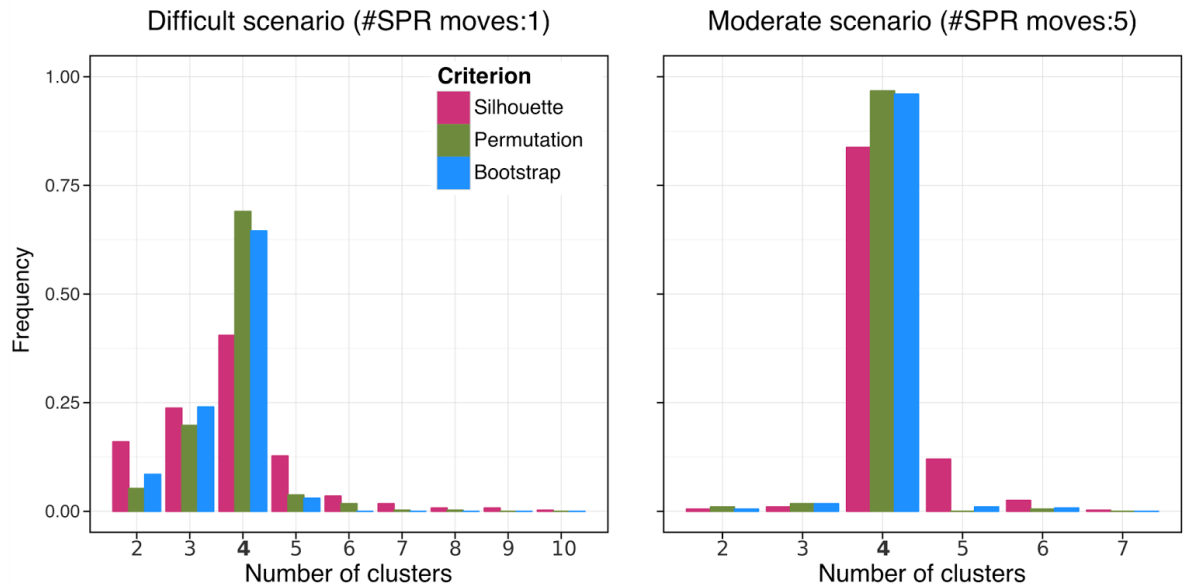

*Figure 4: Aggregate results for 400 "difficult" problem instances (left) and 400 "moderate" instances (right). The true number of clusters is 4. In both sets, our new stopping criteria (permutation and bootstrap) perform better than the general-purpose silhouette method.*

We also considered the performance of the stopping criteria separately for the different distance metrics and clustering methods. In terms of distance metrics, we see little difference between geodesic and Euclidean distances. In contrast, we observe that all three stopping criteria perform noticeably better in combination with spectral clustering than with Ward's method (Supplementary Figure 5). This is particularly the case for our two new criteria (permutation and bootstrap), which outperform the silhouette by a greater margin on the spectral clustering runs.

## Dealing with incomplete occupancy across loci

In the simulations considered so far, we have covered cases in which there has been no missing data. When analysing real data, we cannot guarantee that all loci will be present for all taxa. The effect that missing data have on our method is that we are required to compare trees with different leaf sets, a circumstance for which distance metrics have not been defined. A simple measure to counteract this is to prune trees to the intersection of their taxon sets, and then measure the distance between these reduced trees.

To assess the impact of incomplete occupancy on our approach's ability to infer the correct clusters, we generated additional simulated data sets containing a varying proportion of randomly selected missing genes (see *Materials and Methods*) and analysed the data using the best combination of distance measure and clustering method (geodesic distances and spectral clustering). With missing data, when the number of clusters is known in advance, the true partition of the data is recovered with high accuracy (measured by variation of information) as long as the clusters are separated by a few topological rearrangements—even when data is sparse (Figure 5A). When clusters are not well separated—differing by just 1 or 2 SPRs—sparseness has a detrimental effect on accuracy. Both of the permutation and bootstrap stopping criteria show high accuracy when inferring the number of clusters, strongly outperforming the silhouette (Figure 5B; Supplementary Figure 6).

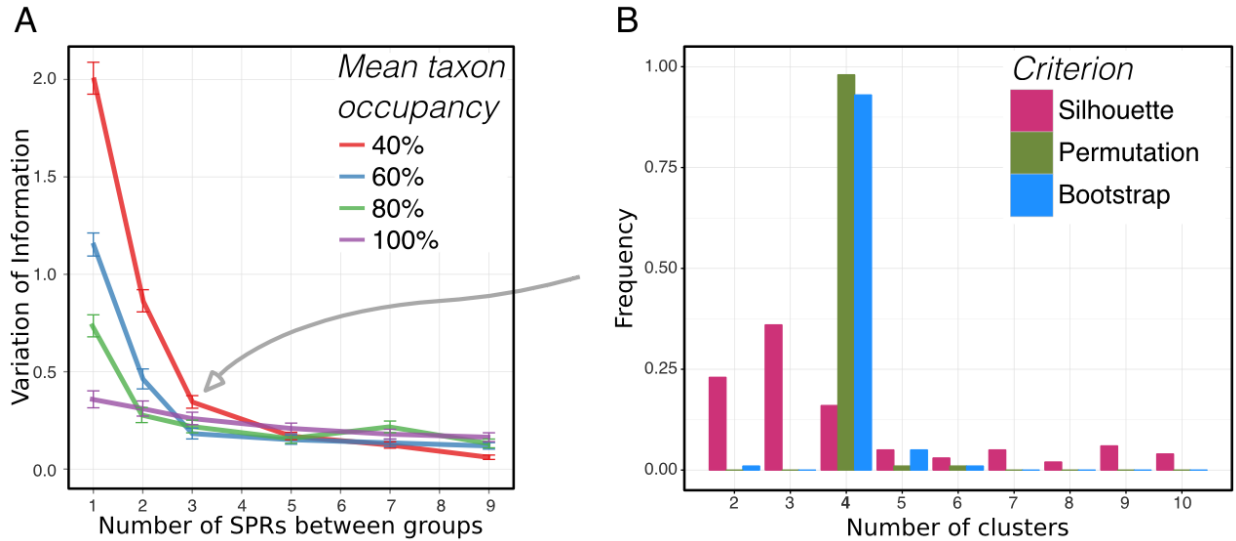

*Figure 5: (A) Distance of the spectral clustering of geodesic distances from the "true" clustering for varying levels of taxon occupancy. Just as with complete groups, partial groups converge to the correct assignment as the distance between clusters increases. When clusters differ from the underlying species tree by 3 SPRs or more, the effect of incomplete occupancy on performance is very slight. (B) Effect of incomplete taxon occupancy on cluster number selection criteria. Non-parametric permutation and parametric bootstrap recover the true number of clusters (4) in more than 90% of cases. The clusters were separated by 3 SPRs, and each locus had 40% mean taxon occupancy, which corresponds to the point on panel (A) indicated by the grey arrow.*

## Application to empirical data

We applied the best combination of distance measure (geodesic distance), clustering method (spectral clustering) and stopping criterion (permutation test) to two empirical data sets.

### *Yeast data set*

The first empirical data set consists of 344 curated orthologous sets of genes from 18 ascomycetous yeast species, which was previously used to infer a species phylogeny robust to inter-gene heterogeneities (Hess and Goldman 2011). Applying our method to this data set resulted in a partition of the 344 loci into three clusters (Supplementary Figure 7). The clusters are of unequal sizes: there is a large cluster, consisting of 307 loci, and two small clusters containing 26 loci and 11 loci. Although the numbering of clusters produced by treeCl has no special meaning, for clarity 'cluster 1' will consistently refer to the cluster of 307 loci, and 'cluster 2' and 'cluster 3' to the clusters of 26 and 11 loci, respectively.

Despite the high degree of incongruence among trees estimated from individual loci, the overall species tree relating these yeasts has been well-studied, and has been established with little controversy (Dujon 2010). This species tree can be seen as the tree on the left in figure 6, which is also the cluster tree derived from cluster 1. The trees on the right of figure 6 are the cluster trees inferred for clusters 2 and 3.

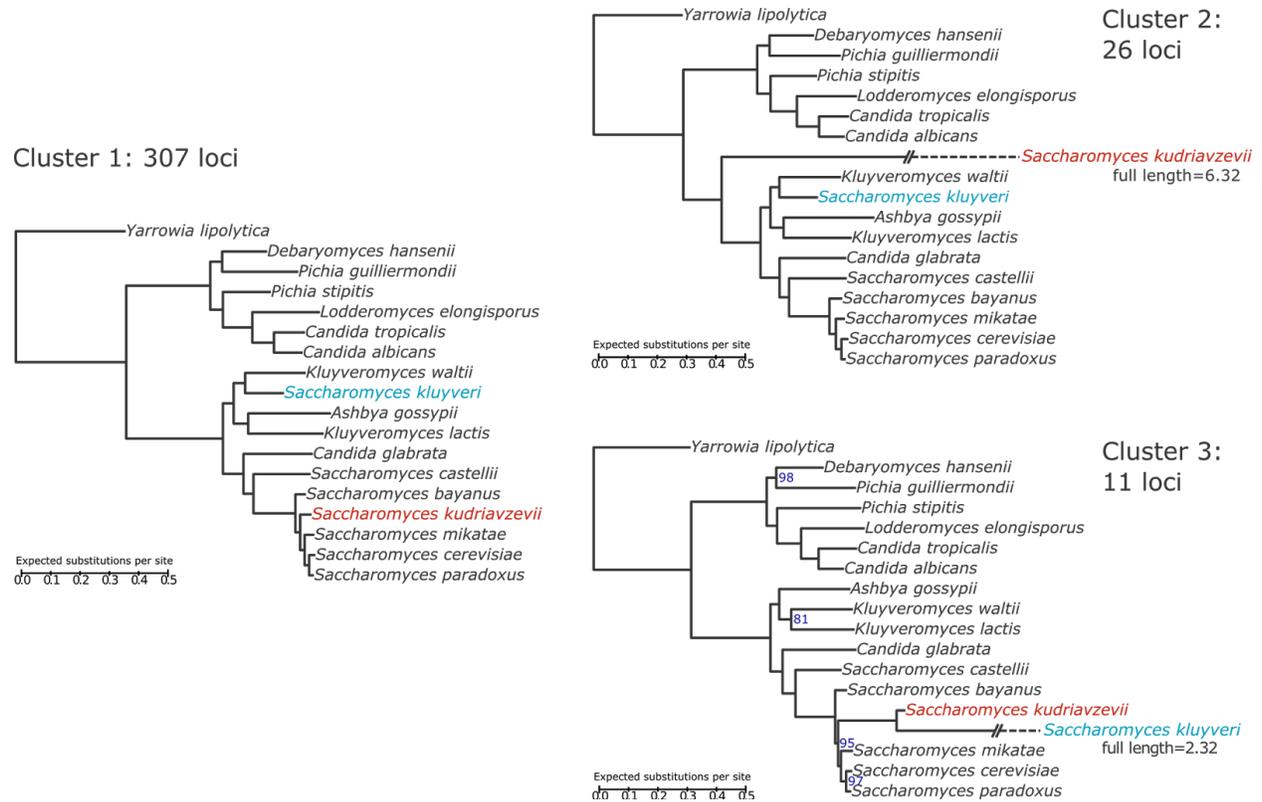

*Figure 6: Phylogenetic trees inferred from the three clusters found in the yeast analysis with treeCl. The tree on the left is that inferred from the largest cluster of 307 loci. This matches the established species tree for these 18 species of yeast. The taxa highlighted in red (*S. kudriavzevii*) and blue (*S. kluyveri*) are those that are found on long branches in the trees inferred from clusters 2 and 3 (shown respectively right, upper, and right, lower). In these trees the branches leading to* S. kudriavzevii *(in cluster 2) and* S. kluyveri *(in cluster 3) have been truncated to so as to fit reasonably on the plot. Their full lengths are as indicated. Otherwise, branch lengths can be determined by the scale bars shown (all equal scales). Branch support measures were calculated using approximate Bayes (aBayes). Where aBayes branch supports are less than the maximum possible value of 100% their values are indicated by a number to the right of the branch.*

The tree for cluster 2 yields nearly the same topology as that for cluster 1, with the sole modification that *S.kudriavzevii* appears basal to, rather than within, the *Saccharomyces sensu stricto* clade. Branch lengths are also modified in the cluster 2 tree: minor changes aside, note that the branch leading to *S. kudriavzevii* is very much longer than in the cluster 1 tree.

A similar observation can be made of the inferred tree from cluster 3. In this case, it is *S. kluyveri* that is incorrectly placed relative to the species tree, again with a very long branch. The cluster 3 tree also differs from the cluster 1 tree in the arrangement of the clade consisting of the species *K. waltii*, *A. gossypii*, *K. lactis*, the clade to which *S. kluyveri* belongs in the other two trees. The cluster 3 tree is also the only one for which the branch support values, as measured using approximate Bayes, are below 100%. The lowest branch support, 81%, is found within the rearranged *K. waltii*, *A. gossypii*, *K. lactis* clade. With this exception, the remaining branches all show greater than 95% approximate Bayes branch support, even though there is incongruence among the loci underlying these trees. However, this may not necessarily be a strong case for these topologies being correct, as it has been suggested that concatenation tends to inflate branch support values (Larget et al. 2010; Weisrock et al. 2012).

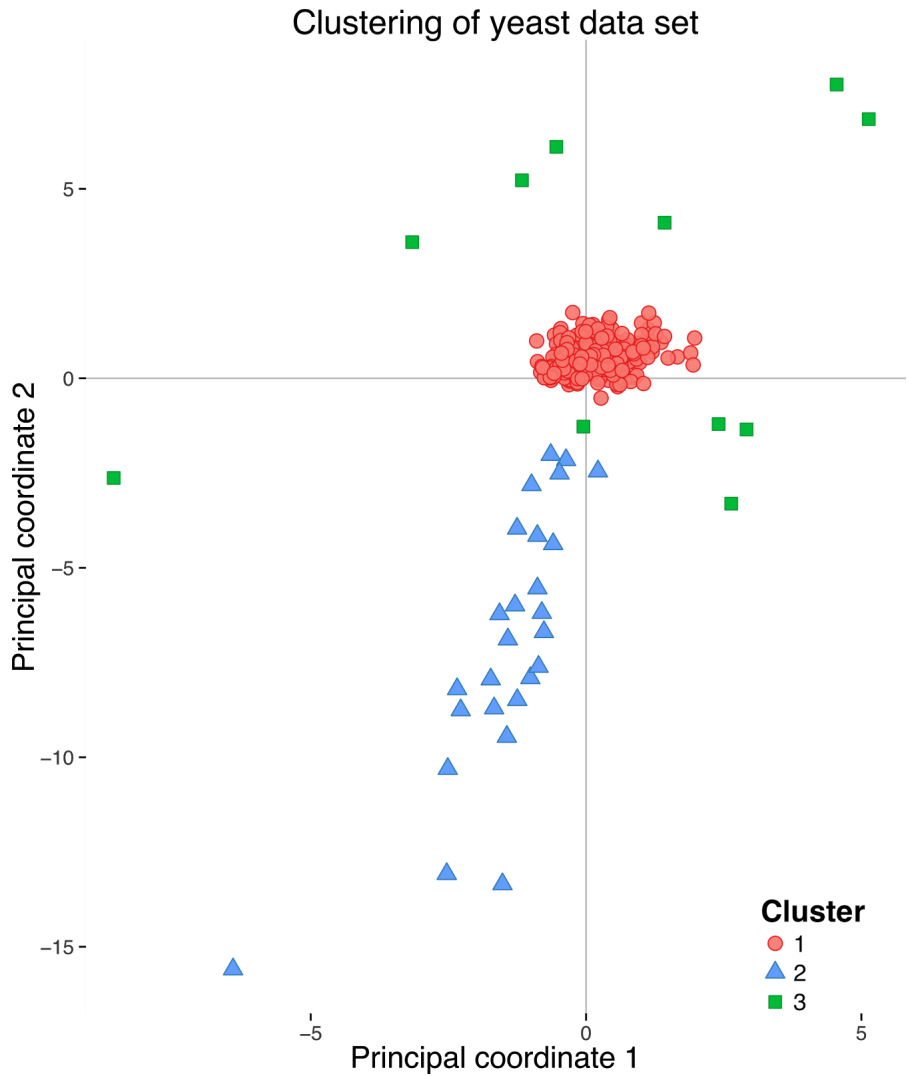

*Figure 7: Visualisation of application of treeCl to the yeast data set. The scatterplot shows the embedding, by multidimensional scaling, of the geodesic distances between the 344 trees. Three clusters were found by spectral clustering: red circles indicate the largest cluster, with 307 members; the 37 remaining loci are indicated by blue triangles (cluster 2) and green squares (cluster 3). Loci belonging to the first, largest cluster are tightly grouped and yield the correct species phylogeny, whereas trees belonging to the second and third clusters are disparate and all have odd and inconsistent phylogenies as a result of incorrectly called orthology (see text for full details).*

As an attempt to visualise the distribution of the individual locus trees we embedded them in two dimensional space using multidimensional scaling (Figure 7). In this representation, cluster 1 appears as a very tight cluster of points in the centre of the figure, while clusters 2 and 3 are more diffuse. Although clusters 1 and 3 appear to overlap, keep in mind that while it may seem to be difficult to assign these clusters on the basis of this figure, the actual clustering is done in a higher dimensional space and using a different coordinate transform than the one visualised here. What can be noted from this figure is that all members of clusters 2 and 3 are positioned relatively large distances away from cluster 1, which suggests that these clusters consist of loci for which the underlying tree distances are large, when measured from those loci from cluster 1.

To try to understand the source of incongruence in the smaller clusters, we examined the sequences associated with the long branch in the gene tree associated with each of their 37 loci. They each included one particularly long terminal branch, but none of the single-locus topologies matched the ones inferred for cluster 2 or cluster 3 as a whole. The 37 trees are reproduced in Supplementary Figure 8. Reciprocal best hit analyses of these sequences with *S. cerevisiae* indicate that they were erroneously classified as orthologs (Supplementary Table 1). We thus conclude that the major source of incongruence in this 344 gene yeast data set is derived from erroneous orthology calling, particularly involving the *S. kudriavzevii* and *S. kluyveri* genomes. In this example, treeCl has identified 307 loci that support the species tree; of 37 that do not, it has detected two clusters, one primarily consisting of cases where the *S. kudriavzevii* gene has been misannotated and one where *S. kluyveri* misannotations are similarly implicated. Even for these two clusters, the inferred phylogeny agrees fully or very nearly with the species tree, aside from the position of the primary misannotated species.

Given the "outlier" nature of the loci identified in the small cluster, we also applied a specialised outlier detection package, kdetrees (Weyenberg et al. 2014). Remarkably, with geodesic distances, it identified the exact same 37 loci as outliers (Supplementary Figure 9). This provides additional evidence that these 37 loci should indeed be excluded in the inference of the species tree.

**Chiastocheta *data set***

The globeflower flies, genus *Chiastocheta,* are pollinators and seed parasites of the plant species from the *Trollius* genus (Ranunculaceae) (Pellmyr 1992; Suchan et al. 2015). *Chiastocheta* have a recent origin, with most diversification events occurring less than ca. 1.6 million years ago, and their phylogenetic relationships are uncertain (Després et al. 2002; Espíndola et al. 2012). Particularly, only two globeflower fly species were found to be phylogenetically supported using mitochondrial markers (Espíndola et al. 2012).

RAD-sequencing of 306 samples from 7 European *Chiastocheta* species (25 *C. dentifera* individuals, 48 *C. inermella,* 52 *C. lophota,* 34 *C. macropyga,* 70 *C. rotundiventris, 36 C. setifera, 41 C. trollii*) collected across their whole ranges yielded a data matrix of 5574 orthologous sets of sequences (loci), containing in total 253866 variable, and 81379 parsimony informative sites. Because of inherent technical limitations of RAD-sequencing, the majority of these loci had sparse coverage over the individuals. To focus on the phylogenetically most informative loci, we disregarded loci present in fewer than 100 individuals. This resulted in a matrix of 176 loci (i.e., 10.2% of the overall number of loci identified). Each locus contained, on average, 44.2% of the taxon set.

Application of treeCl (with geodesic distance, spectral clustering, and permutation test stopping criterion) identified eight clusters. However, the plot of the likelihood improvement against the number of clusters (Figure 8) is not smooth: most of the improvement is obtained by increasing the number of clusters up to four and by increasing it from five to six; in contrast, adding a fifth or seventh cluster only moderately improves the fit. Thus, a cautious interpretation of this analysis is that there are at least four distinct clusters of loci. This conclusion is also supported by the parametric bootstrap criterion (Supplementary Figure 10).

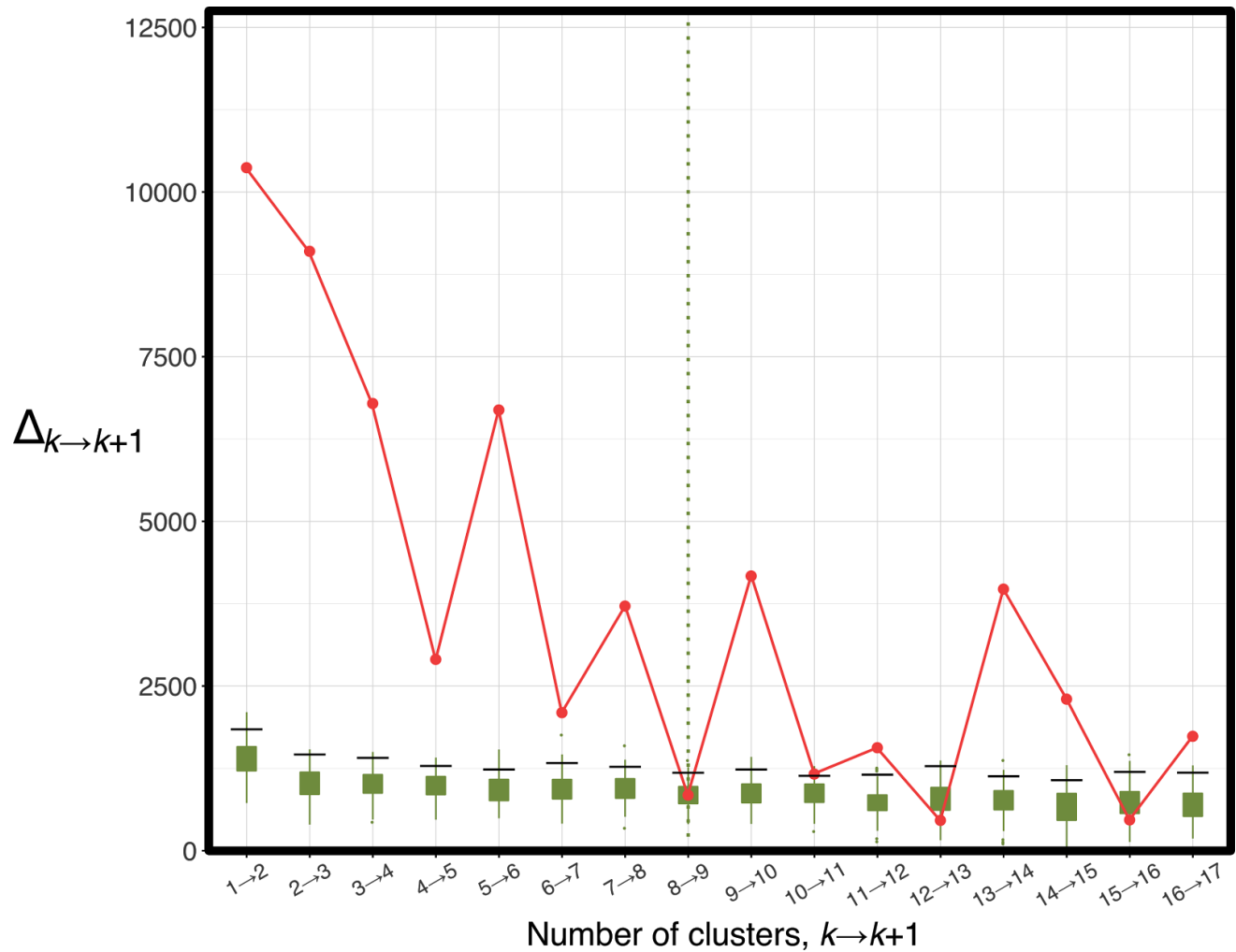

Figure 8: Likelihood improvement gained when partitioning the Chiastocheta data into increasing numbers of clusters (red points). Resampled distributions (boxplots) were generated using the permutation procedure. The number of clusters selected by the stopping criterion is indicated by the vertical dashed line. For 2–8 clusters the improvement is statistically significant; increasing to 9 clusters is not.

The trees inferred for the four clusters (Figure 9) substantially differ, both in topology and branch lengths. In particular, many of the deep relationships are well-resolved but different across clusters, suggesting genuine differences in the history of the loci. However, with very few exceptions, each species forms a distinct monophyletic group. This is consistent with well-documented differences in genital morphology across most of these species (Després et al. 2002). With greater data available, phylogeny and morphology now agree. Furthermore, the even cluster size distribution (cluster sizes of 29, 58, 42 and 47 loci) suggests that the method is not simply finding groups that consist of one or two outliers. The greatest departure from monophyly is shown in the group consisting of 29 loci. In this group the majority of the representatives of species *C. lophota* are found at the base of a clade that also contains *C. macropyga*, *C. trollii*, *C. setifera* and *C. inermella*. For partitions into greater numbers of clusters than four, we observe at least one tree in which species monophyly is largely absent (Supplementary Figure 11), which may indicate that likelihood improvements gained when clustering into more than four groups are due to fitting to the noise in the data, extracting loci with weak or conflicting signal. In this case, attempting to visualise the individual locus trees in two dimensional space does not yield informative results (Supplementary Figure 12).

Overall, the picture that emerges from the analysis confirms the existence of seven distinct species in the *Chiastocheta* genus, but implies that the branching order among them varies substantially across loci. Such variation is suggestive of ILS, particularly as six of the seven species (except *C. rotundiventris*) are thought to have radiated more or less synchronously (Espíndola et al. 2012). To rigorously test this hypothesis, future work could assess the fit of this data under a mechanistic model of ILS.

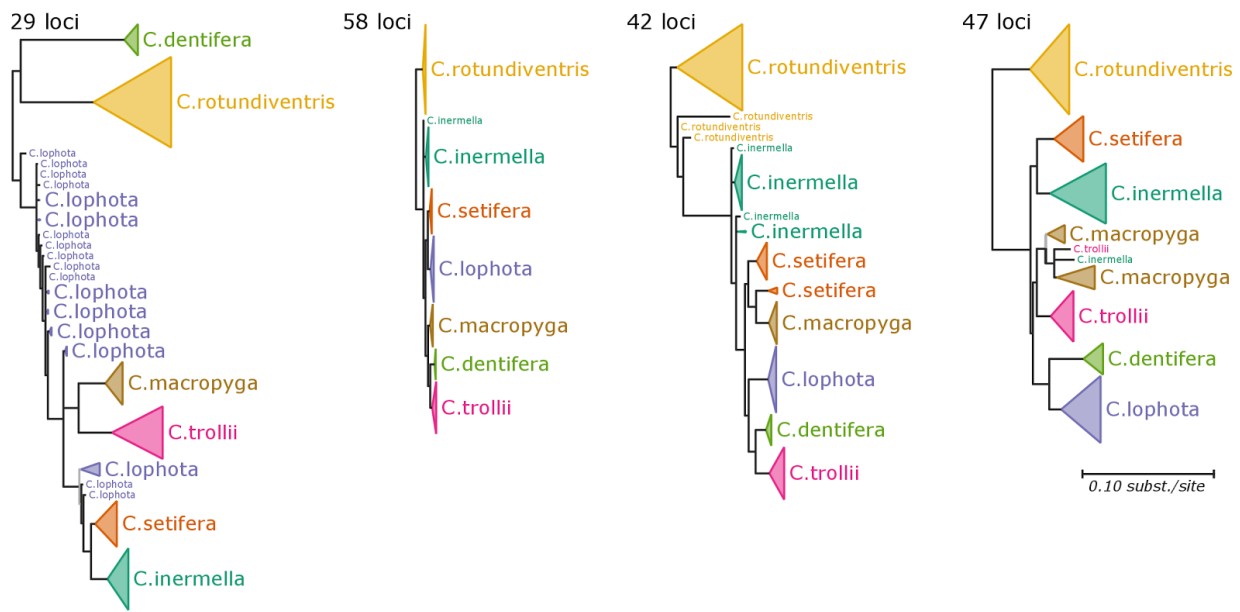

*Figure 9: Trees obtained when clustering RAD-seq data from globeflower flies of the genus* Chiastocheta. *The trees are drawn to scale, and are rooted at their midpoint, as the outgroup is unknown. Leaves are coloured according to species membership. Branch support is indicated as follows: branches with support values below 0.9 are collapsed into multifurcations; those with support in the range 0.9–0.95 are coloured grey; those with support >0.95 are coloured black. Support values are calculated using approximate Bayes (Anisimova et al. 2011).*

## Discussion

In this study, we investigated clustering multi-locus data sets into evolutionarily similar groups based on their inferred phylogenies. This work is motivated by the observation that phylogenetic incongruence among loci can arise through various evolutionary processes, in which case a single tree is insufficient to describe the disparate processes underlying the data. At the other extreme, reporting one tree for each locus suffers from the drawbacks of single-locus phylogenetics—lack of signal, sampling error, unrepresentativeness—and in addition it is difficult to interpret a large and unwieldy collection of trees. By clustering loci, we allow the possibility that a meaningful representation be given by some intermediate number of trees, each capturing a common evolutionary history for some of the loci. We do this in a process-agnostic way, in that we do not seek to view our observations through the lens of any particular mechanism. This may lose inferential power in the case where organisms have evolved mainly through a process that we fail to model explicitly, but has the advantage that we will not bias the analysis by imposing mathematical models inappropriate for the processes that have occurred.

To investigate the performance of this approach, we assessed combinations of different distance metrics and clustering methods using simulation. Overall, Euclidean and geodesic distances, which take branch lengths into account, performed better than Robinson-Foulds distances. Spectral clustering and Ward's method gave the best clusters, most reliably over the range of simulations analysed. We note that two methods—MDS and *k*-medoids—are successful in many cases, but produce some anomalous results in which performance becomes worse as the problems become easier (Supplementary Figures 1D, 2A, 3E).

We introduced new statistical tests to determine the best-supported number of clusters, and compared them to a general-purpose cluster assessment statistic. In simulation the new measures outperformed the general-purpose criterion. If we look at the results from the "difficult" case, it seems that all criteria have a tendency to be conservative and underestimate the number of clusters (Figure 4, Supplementary Figure 5). We consider this a valuable feature; it is more parsimonious to erroneously infer too few rather than too many clusters. When moving from simulated data to a real data set of 344 orthologous groups from yeast (Hess and Goldman 2011), subtle errors in orthology inference could be detected and corrected. This highlights the high potential of the approach for quality control in multilocus phylogenetic analyses. Furthermore, the unexpected nature of the errors observed in that data set is a good illustration of the flexibility of process-agnostic methods for detecting incongruence. The examination of a large data set of *Chiastocheta* flies demonstrates that our method is applicable to data sets of the scale that is routinely produced by high-throughput sequencing approaches such as RAD-seq, and not only to more artificial simulations.

The range of methods and conditions investigated in this study is considerable, but inevitably not exhaustive. There are other distance metrics and clustering methods not tested here. These were omitted mainly for reasons of being too numerous for their inclusion to be practical. Some were not considered because they overlapped closely with metrics and methods that were considered: for instance, kernel PCA is a coordinate transformation procedure that could have been used in a similar way to spectral embedding and multidimensional scaling; however, it is largely analogous to spectral embedding (Ng et al. 2002) and initial investigation showed it to give very similar results. Other clustering methods such as Markov Clustering (Enright et al. 2002), DBScan (Ester et al. 1996) and Affinity Propagation (Frey and Dueck 2007) were not investigated as they provide no means to specify the number of clusters they return, which is a property we specifically wanted so we could test our stopping criteria. Similar concerns led to us to exclude such distance measures as Quartet Distance (Estabrook et al. 1985) or Matching (Lin et al. 2012) as they provide discrete topology-only measures similar to Robinson-Foulds. Tree edit measures such as the subtree prune-and-regraft distance are highly computationally difficult to calculate (Bordewich and Semple 2005) and so were not investigated. This method may become tractable with the advent of fast approximation algorithms (Chung et al. 2013; Whidden et al. 2013).

We tested our method under a range of simulation criteria. However, the combinatorics of the range of parameters that can be varied are such that it was not possible to test them all. This also limits the degree to which we can test whether tuning certain clustering procedures might improve their performance (for instance, the number of dimensions to embed the inter-tree distances in when using MDS). Likewise, many biological phenomena leading to incongruence were not investigated (including variation in rate of evolution across genes and between taxa; differential duplication and loss between species within gene families; etc.). Nevertheless, we think that the variety of problems studied, and in particular the range of levels of difficulty, are enough to provide convincing evidence that process-agnostic clustering methods can work effectively and give useful results.

The clustering methods investigated in this work are also applicable to data sets with incomplete

"occupancy" among species, such as the one obtained for *Chiastocheta* flies by RAD sequencing, a technique that is typically prone to having a large proportion of missing data. Indeed, our simulations suggest that as long as the clusters are separated by a few topological moves, occupancy as low as 40% incurs negligible performance degradation. Likewise, the new stopping criteria introduced in this study cope well with sparse data matrices, in contrast to the general-purpose silhouette method.

It is unclear how sensitive the method is to the quality of the inferred single-locus trees. Inferring these is the first step in our analyses, and all further steps proceed as if the trees are correct; the distance matrix is calculated based on these initial trees, which are not re-estimated. To improve our approach we could introduce a cycle in our algorithm in which the single-locus trees are re-estimated based on parameters estimated while inferring the cluster trees, and the distance matrix and cluster assignments updated. However, this is likely to be computationally expensive. Another possibility is to incorporate measures of phylogenetic uncertainty—such as the bootstrap—into the distance estimation and the clustering step.

Practically, however, the amount of computation required to apply distance metrics and clustering methods to whole-genome-scale data poses a challenge. For instance, calculating geodesic distances takes time of order $O(n^4)$ (Owen and Provan 2011), where $n$ is the number of leaves in the tree, while Euclidean and Robinson-Foulds distances can be computed in linear time (Pattengale et al. 2007). There is also the burden of pruning trees to their overlapping taxa. These factors could prove prohibitive in the case of very large trees. Whatever the details of the distance calculations, they must be performed $\binom{m}{2}$ times, where $m$ is the number of loci in the data set. Clustering the resulting $m \times m$ distance matrix using any spectral technique—requiring eigen decomposition—takes time of order $O(m^3)$. This burden can be reduced by applying an approximation such as the Nyström method (Fowlkes et al. 2004), which produces approximations to the eigenvalues and eigenvectors from a reduced input set, reducing the number of pairwise tree distance comparisons required. We have demonstrated that the relatively efficient Euclidean distance and Ward's method for hierarchical clustering produce good results, and may thus be preferred in large data sets. In the work carried out in this paper, by far the largest amount of time is spent in tree inference; this remains the bottleneck.

We applied our method to two empirical data sets, one from yeasts and one from *Chiastocheta* flies. Both data sets show a high degree of phylogenetic incongruence, although this is likely to be for different reasons: misannotated orthology for the yeast data set, and ILS for *Chiastocheta*. Due to its process-agnostic nature, we were able to apply our method in the same way to both data sets, and learn something about the incongruent signals in the data. This allows us to identify the likely processes at play, and prioritise different types of follow-up analysis—stringent orthology identification in the first case, and analysis under a mechanistic ILS model in the second. In this way our process-agnostic is complementary, rather than in opposition, to mechanistic models of incongruence.

Looking ahead, it seems clear that the assumption in multi-locus phylogenetics that all loci are derived from the same tree is too strong, and should be relaxed. Partitioning model parameters is commonplace (e.g. Hess and Goldman 2011; Lanfear et al. 2012); tree-topology partitioning is a logical next step.

# Materials and Methods

In the following subsections, we first describe the components of this clustering process in more detail, including the various distance and clustering algorithms investigated in this study. Next, we describe a partition likelihood quality score that we use to compare the performance of combinations of distances and clustering methods, and introduce new tests to infer the optimal number of clusters in a data set. Finally, we describe the simulated and empirical data used in our analyses. The analyses were carried out using our treeCl software, which is available as an open source python package (http://git.io/treeCl).

### Input data

The input data are a set of multiple sequence alignments, one per locus being examined. The sequences can be of nucleotides or proteins.

### Tree inference

In principle, any method of tree estimation can be used. We use maximum likelihood (ML) estimation of phylogenies, which is statistically robust (Felsenstein 2004) and enables us to use a likelihood criterion for cluster membership comparisons and cluster number decisions. For each locus, we infer the ML phylogenetic tree using the Phylogenetic Likelihood Library (PLL) (Flouri et al. 2015).

In the experiments described in this paper, we use PLL's full ML estimation with tree search. We use either the GTR model (Tavaré 1986) for nucleotide data, or the WAG model (Whelan and Goldman 2001) for proteins, coupled with a gamma distributed model of rate variation with 4 discrete categories (Yang 1994), and the RAxML search strategy (Stamatakis 2014).

## Inter-tree distances

Once the tree for each locus has been estimated, their similarities are assessed according to a particular distance metric. We have investigated three distance measures: Robinson-Foulds (Robinson and Foulds 1981), Euclidean (Kuhner and Felsenstein 1994) and geodesic (Billera et al. 2001) (table 1). With a set of *m* trees we compute all *m(m*-1)/2 pairwise distances. We implemented the tree distance algorithms in C++ and Python. The geodesic distance algorithm used is that of Owen and Provan (2011). Source code is available from https://pypi.python.org/pypi/tree_distance/0.0.6.

| Distance measure | Features incorporated |
| --- | --- |
| Robinson-Foulds | Topology |
| Euclidean | Branch lengths |
| Geodesic | Topology and branch lengths |

*Table 1: Distance metrics investigated*

### *Missing data*

For pairwise tree comparisons when taxon sets differ, the trees are pruned to the taxa they have in common. Distances are calculated on the resulting reduced trees. In the case that the intersection of taxon sets contains fewer than four taxa—the minimum number required that can produce a tree with at least one internal edge—the distance is taken to be zero.

## Clustering

The resulting distance matrix is used as the input for a clustering algorithm. We have investigated seven such algorithms, detailed in table 2. Each algorithm presumes that the required number of clusters is known in advance; we investigate approaches for choosing the optimal number of clusters below. All methods work directly on the distance matrix, except the coordinate transform methods. These transform the distance matrix into the coordinates of a set of points, then use *k*-means to perform the final clustering step. *k*-means is not suitable for use directly on a distance matrix.

| Clustering method | Type | Implementation |
| --- | --- | --- |

| Single-linkage | Hierarchical | Fastcluster (Müllner 2013) |
|---|---|---|
| Complete-linkage | Hierarchical | Fastcluster |
| Average-linkage (UPGMA) | Hierarchical | Fastcluster |
| Ward's method | Hierarchical | Fastcluster |
| Spectral Clustering (using *k*-means for the final clustering step) | Coordinate transform | Spectral clustering: custom implementation in treeCl (after Zelnik-Manor and Perona 2004) *k*-means: Scikit-learn (Pedregosa et al. 2011) |
| Multidimensional scaling (MDS) + *k*-means | Coordinate transform | Custom implementation in treeCl (after Torgerson 1952) |
| *k*-medoids | Partitioning around medoids | C Clustering Library (de Hoon et al. 2004) |

*Table 2: Clustering methods investigated*

## Partition likelihood for assessing clustering

In order to assess partitions, which may be obtained from different clustering approaches, we describe the 'partition likelihood'. This can be used as a quality score to assess the best combination of distance and clustering method.

Each cluster comprises a subset of the loci, and is a collection of genes putatively sharing a common evolutionary history. Hoping to benefit from a more robust evolutionary inference by combining the data from homogeneous sources, we therefore concatenate the alignments of the member loci and infer the ML tree using the same model as for the individual loci. The log-likelihood is calculated for each cluster tree conditioned on the concatenated cluster alignment. The partition log-likelihood, $\ell^P$, is the sum of all optimal cluster log-likelihoods, and is in effect the maximum log-likelihood under a model where the genes within each cluster share a common evolutionary history and evolutionary dynamics, but there are no constraints that different clusters share any evolutionary parameters.

## Choice of number of clusters

The number of clusters, *k*, can take any integer value in the interval [1, *m*], where *m* is the number of loci in the data set. Let us consider the case of choosing between *k* and *k*+1 clusters. This is equivalent to choosing between the hypotheses that the loci are sampled from *k* evolutionary trees, or *k*+1 evolutionary trees. These form our null and alternative hypotheses, respectively. The alternative hypothesis is able to recapitulate the null model, and therefore the hypotheses are nested. To illustrate that the alternative hypothesis nests the null, consider that if two of the trees associated with clusters in the alternative model are identical it is equivalent to the case that those clusters are combined, decreasing the effective number of clusters by one and reproducing the null. We can thus calculate the partition log-likelihood of each hypothesis, $\ell^P_k$ and $\ell^P_{k+1}$, and the increase in log-likelihood, $\Delta_k = \ell^P_{k+1} - \ell^P_k$.

With nested hypotheses, $2\Delta_k$ is asymptotically chi-squared-distributed, with the number of degrees of freedom corresponding to the difference in the number of parameters between the null and alternative hypotheses (Wilks 1938). However, counting parameters proves difficult in this case: the extra parameters in the alternative hypothesis include an inferred tree topology, and tree topology parameters are difficult to quantify (Goldman 1993). This means we cannot specify which chi-squared distribution we should use for our test. This also precludes the application of information criteria such as the AIC or BIC (Akaike 1974; Schwarz 1978).

Alternatively, we can estimate the distribution of $\Delta_k$ by repeatedly calculating $\Delta_k$ values from new data sets generated under the null hypothesis. Such a procedure does not require that the difference in degrees of freedom be known or even that the hypotheses be nested. We devised two such procedures: a non-parametric permutation test, in which the new data sets are produced by randomising the original data, and a parametric bootstrap test, in which new data sets are generated via simulation. These permit us to compare whether $k$+1 clusters are statistically supported over $k$ clusters, and we apply such tests successively for $k$ = 1, 2, 3, … and use the stopping criterion that $k$* clusters are taken to be optimal where $k$* is the smallest value of $k$ for which $k$+1 clusters are not statistically supported over $k$ clusters.

*Permutation test*

The permutation test generates a new data set from the input data set by permuting the columns of all the multiple sequence alignments—the alignments are concatenated, the columns are shuffled, and the concatenated alignment is broken back up into individual alignments of the same lengths as the original ones. The effect of this is to uniformly distribute the columns over the data set, removing any between-locus incongruence that might form the basis for clustering. These resampled data are analysed twice: into $k$ clusters and $k$+1, and we calculate $\Delta_k$. The whole permutation procedure is repeated 100 times to estimate the distribution of $\Delta_k$.

Note that it would be conceptually preferable to permute the columns such that a distribution of loci among exactly $k$ underlying trees is preserved (as per the null hypothesis). However, we have not found a good way to do so. Thus, we implicitly assume that the distribution of the improvement in likelihood from $k$ to $k$+1 is the same whether the true number of clusters is 1 or $k$. Our extensive simulations suggest that this approximation works well in practice.

*Parametric bootstrap*

As a parametric alternative to the non-parametric permutation test, we use simulation to generate new data sets using parameters estimated during the analysis of the original data. After the analysis, each locus belongs to one of $k$ clusters, and is therefore associated with one of $k$ cluster trees. In the simulated data set, each locus is simulated along its associated cluster tree, using evolutionary model parameters estimated in the analysis. Alignment length and gap positions are duplicated from the initial data (Goldman et al. 1998). Consequently, the data is simulated under the null hypothesis that loci evolved along $k$ underlying trees. The simulated data are clustered and separately analysed with $k$ clusters and with $k$+1 clusters to calculate the increase in the partition log-likelihood, $\Delta_k$. This "parametric bootstrap" procedure is repeated for 100 data sets to estimate the distribution of $\Delta_k$. The simulation code makes use of the Bio++ libraries (Guéguen et al. 2013).

## Simulating data sets with incongruence

The simulated data used in this study were generated to represent evolutionary histories with incongruent phylogenies. Consequently, generating the simulated data involved three stages: (i) deciding on the number of taxa, clusters, loci, and distribution of loci into clusters; (ii) for each cluster, generating an evolutionary tree; (iii) for each locus, simulating sequences along its cluster's tree.

### *Number of taxa, clusters, loci, and distribution of loci into clusters*

We produced data sets according to four scenarios with varying numbers of taxa, loci, clusters, and distribution of loci among these clusters, as described in table 3.

| Name | Taxa | Clusters | Loci | Distribution of loci into clusters |
|---|---|---|---|---|
| Small uniform | 20 | 4 | 60 | 15, 15, 15, 15 |
| Small skewed | 20 | 4 | 60 | 5, 10, 15, 30 |
| Large uniform | 40 | 6 | 90 | 15, 15, 15, 15, 15, 15 |
| Large skewed | 40 | 6 | 90 | 5, 5, 10, 10, 20, 40 |
| Incomplete occupancy | 50 | 4 | 60 | 15, 15, 15, 15 |

*Table 3: Attributes of the four simulated data set scenarios with incongruence used to test combinations of distance metric and clustering method, and the scenario used to test the effect of incomplete occupancy.*

### *Generating cluster trees*

All cluster trees are derived from an underlying 'species tree'. For each data set, we simulated a random species tree using a Yule pure speciation model (Yule 1925), implemented in Dendropy (Sukumaran and Holder 2010).

To generate incongruent cluster trees, we started from this species tree and applied sequences of random rearrangements of potential biological relevance. The type of rearrangements was either nearest-neighbour interchange (NNI) or subtree prune-and-regraft (SPR). NNI makes local rearrangements, such as those that might be found as a result of ILS. SPRs were used to make rearrangements involving branches at a greater separation within the tree, consistent with the kind of rearrangements observed in HGT (e.g. Galtier 2007).

We applied a predetermined number of rearrangements to the underlying tree for any given data set. This number was varied to control the 'difficulty' of the data set, i.e. the expected difficulty for a clustering method to reproduce the correct partition of the data. A data set with a small number of rearrangements is derived from cluster trees that are more similar to each other than one with a large number of rearrangements, and therefore represents a more difficult case. The number of rearrangements we used ranged from 1 to 10; beyond 10 NNIs or SPRs the underlying trees were so different that all clustering strategies performed so well that there was no distinction between them.

Combining the four scenarios from the previous section with the two rearrangement types and ten difficulty levels yields 80 different parameterisations that describe the attributes of the data sets we generate.

### *Simulating data sets for testing combinations*

For each parameterisation we generated 1000 replicate data sets according to the following process:

I. Randomly generate an ultrametric species tree according to the Yule process.
II. For each cluster, apply a sequence of random tree rearrangements to the species tree to generate the cluster tree. The species tree is reset at the end of the sequence of rearrangements, so that it is identical for each cluster prior to the rearrangements being applied. The rearrangements are either NNI or SPR. The branches at which these operations are applied are selected randomly according to the following procedure: the tree length, *L*, is the sum of all branch lengths. A line (0, L) can be interpreted as all the branches in the tree laid end-to-end. A random value drawn from U(0, L) gives us both a randomly selected branch—according to the branch segment it falls in—and a position on that branch.
III. Draw a set of branch lengths for the cluster trees: inner branch lengths are set to values drawn from Gamma(*shape*=0.67, *scale*=0.16), terminal branch lengths to values drawn from Gamma(*shape*=0.54, *scale*=0.48). These distributions were fit to the branch lengths inferred for the yeast data set.

IV. Simulate alignments from each cluster tree according to the distribution of loci into clusters. Protein sequences were simulated using ALF (Dalquen et al. 2012), using the WAG model of substitution (Whelan and Goldman 2001) with 4 categories of gamma distributed rates ($\alpha$=1); (Yang 1994). Sequence lengths were drawn from a gamma distribution with shape=1.772 and scale=279.9. These parameters were estimated from the distribution of alignment lengths of the yeast data set (see section Empirical Data, Yeasts).

V. Sequences were removed from the alignments with probability (1 - occupancy).

## Empirical data

**Yeasts**

After validating the performance of our method under the controlled conditions of simulation, we investigated its performance on a data set of 344 orthologous groups from 18 yeast species (Hess and Goldman 2011). We analysed protein sequences using the WAG model (Whelan and Goldman 2001). The loci were clustered based on geodesic distances and spectral clustering, with the number of clusters determined by parametric bootstrap.

*Chiastocheta*

The second data set consisted of the RAD sequences obtained from *Chiastocheta* flies (Diptera: Anthomyiidae) collected across their whole European range. Samples were genotyped using a modified ddRAD protocol (Peterson et al. 2012; Mastretta-Yanes et al. 2015). *De novo* locus assembly was performed using the pyRAD 2.0 package (Eaton 2014), with read clustering similarity threshold of 75%, both on within- and among-sample level. Other parameters were set as follows: all nucleotides with Phred quality lower than 20 were treated as unknown bases, and reads with more than 4 unknown bases were removed from the data set; possible paralogs were removed by filtering out the loci that had more than five heterozygous positions per locus within individuals, more than 10 heterozygotes per nucleotide position among samples, and the loci for which more than two alleles were present per individual. In total 273 individuals were sequenced, with 33 technical replicates. For the purpose of this study, only high coverage loci (i.e. present in at least 100 samples) were retained. This resulted in a matrix of 176 loci across 306 samples. Phylogenetic analysis was performed using the GTR model + 4 categories of Gamma distributed rates across sites. Clustering parameters were Geodesic distances, spectral clustering, and the number of clusters was estimated using the non-parametric permutation test stopping criterion.

## Data available for download

Simulation data and results from ms. sections 'Performance of the combinations of distance metrics and clustering methods', 'Performance of methods for determining the number of clusters' and 'Dealing with incomplete occupancy across loci', and the alignments and trees for the original loci and for the optimal clusters for the yeast (344 loci; 3 clusters) and *Chiastocheta* (176 loci; 4 clusters) data sets, are available for download from http://www.ebi.ac.uk/goldman-srv/treeCl.

## Acknowledgements

KG and NG acknowledge funding from the European Molecular Biology Laboratory. The work was also supported by Swiss National Science Foundation grants PP00P3_150654 to CD and PP00P3_144870 to NA.

# Supplementary Figures and Tables

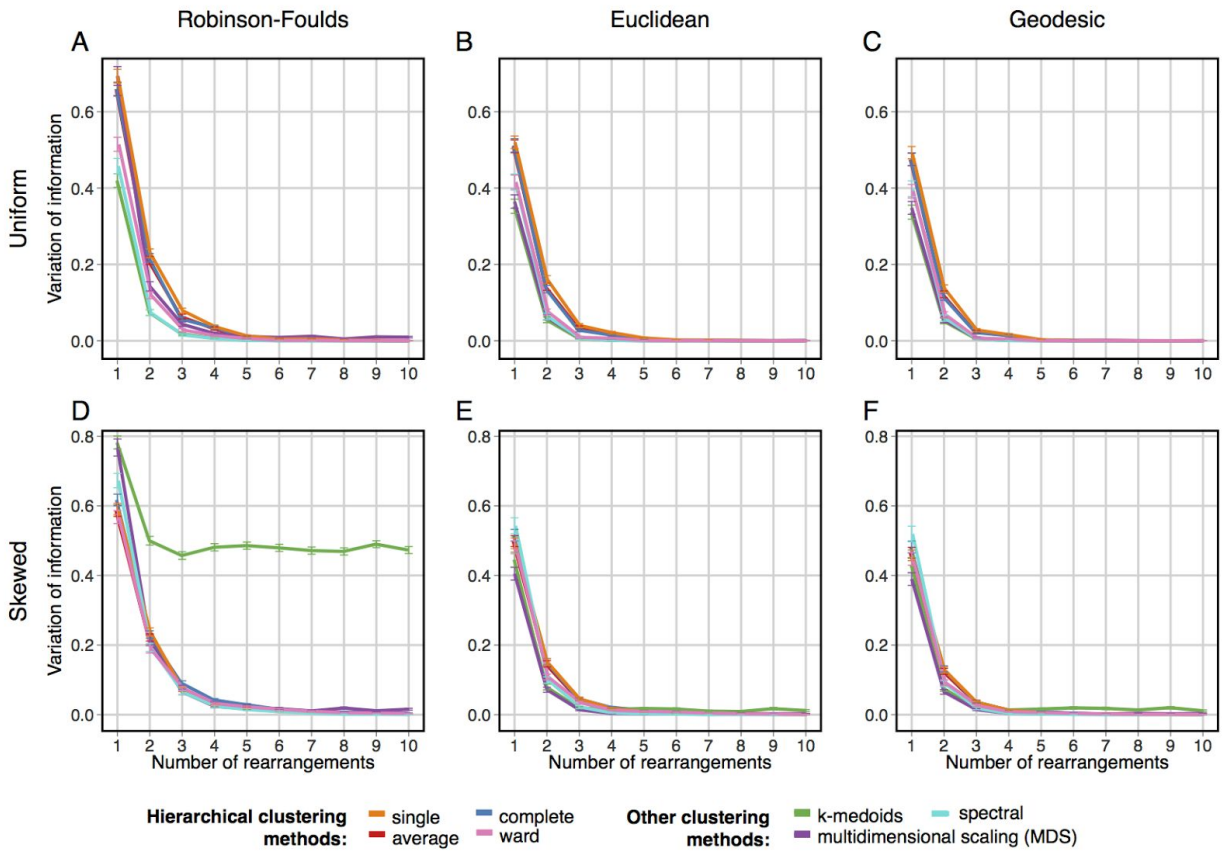

*Supplementary Figure 1:* **Small data set, SPR rearrangements.** *Panels show the relative performances of combinations of distance metric (varying over columns of panels) and clustering methods (shown by the colours of the lines), as measured by the variation of information metric (y-axes), which is a measure obtained when comparing the inferred solution with the true solution (higher values show a larger departure from the correct solution). Lines show the mean value obtained from 1000 replicates, and the error bars show the standard error of the mean. Rows correspond to the experiments with a partition of uniformly-sized clusters (A–C) and those with a partition of clusters of skewed sizes (D–F). In each individual panel, the x-axis represents the number of SPR rearrangements separating the underlying clusters, so that increasing values along this axis correlate with the clustering problem becoming easier.*

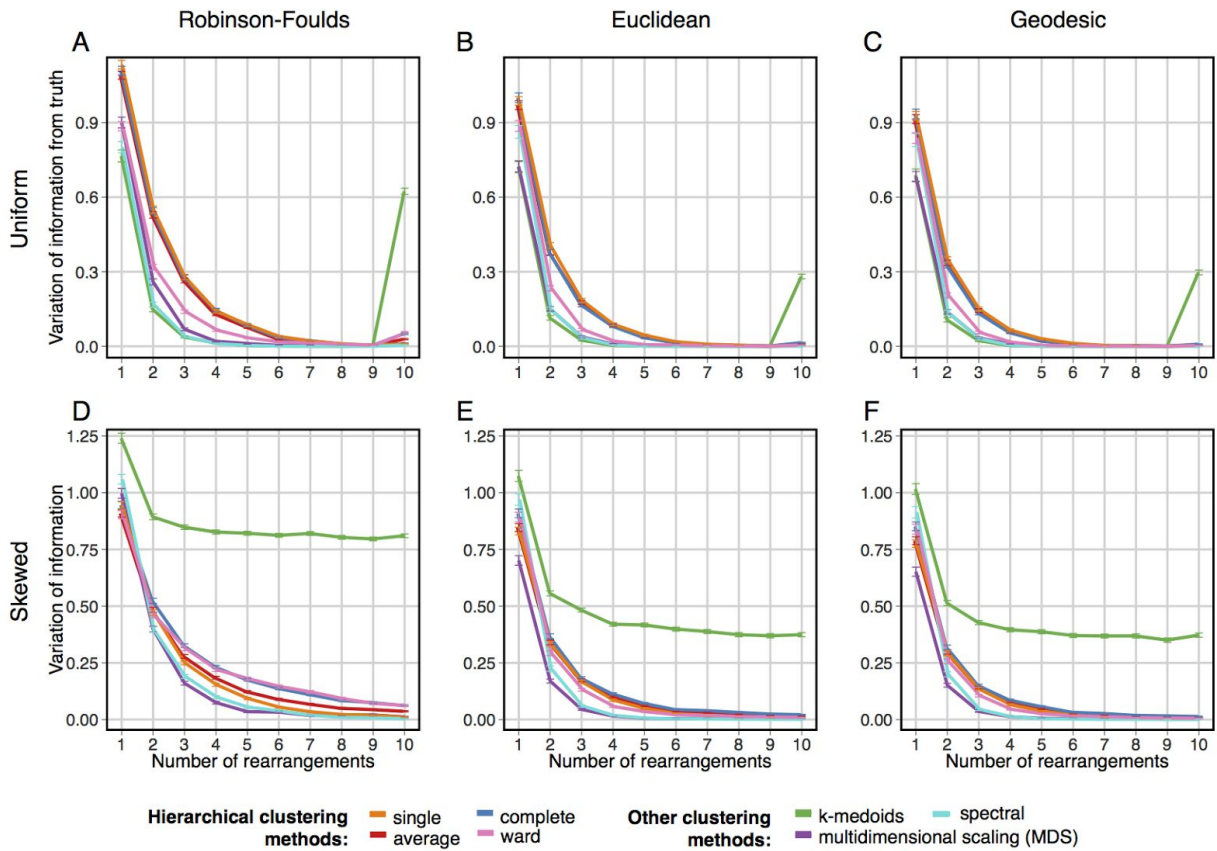

*Supplementary Figure 2:* **Large data set, SPR rearrangements.** *Panels show the relative performances of combinations of distance metric (varying over columns of panels) and clustering methods (shown by the colours of the lines), as measured by the variation of information metric (y-axes), which is a measure obtained when comparing the inferred solution with the true solution (higher values show a larger departure from the correct solution). In each individual panel, the x-axis represents the number of SPR rearrangements separating the underlying clusters.*

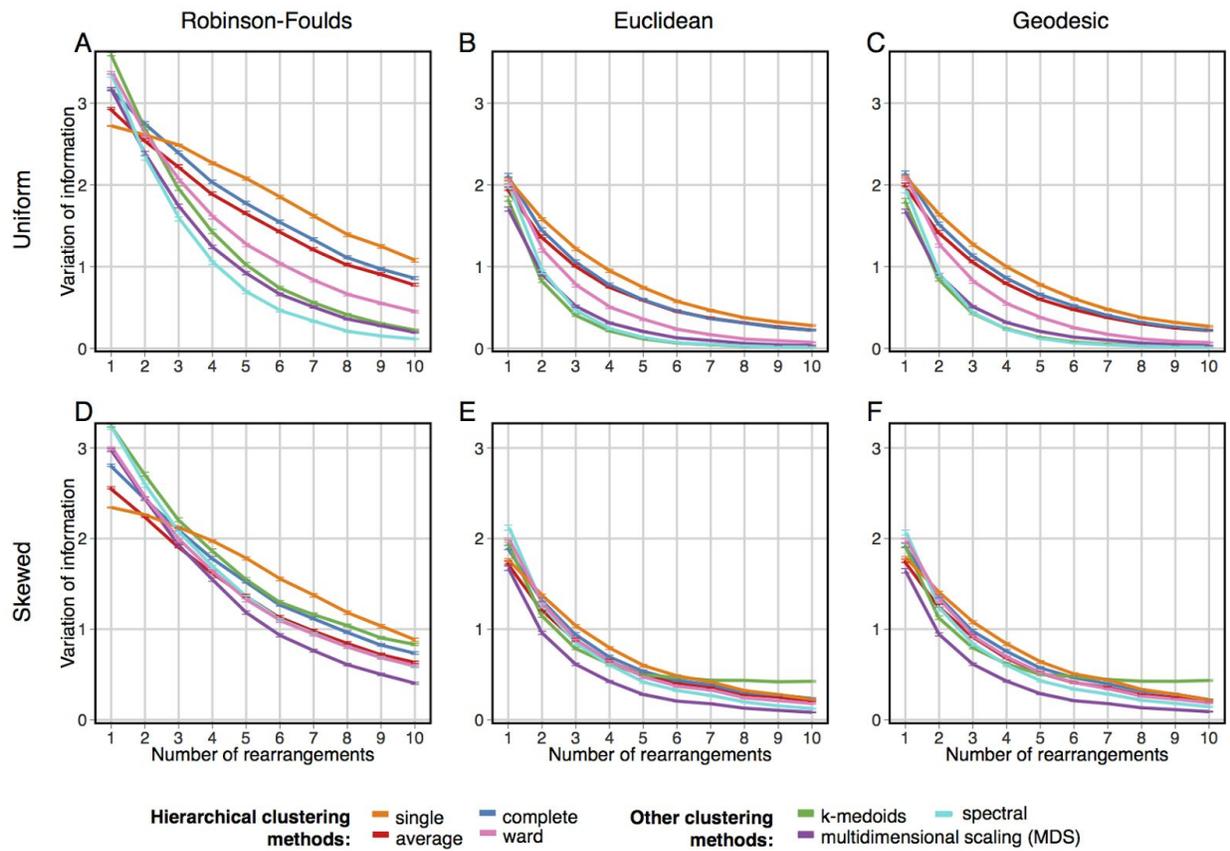

*Supplementary Figure 3: **Large data set, NNI rearrangements**. Panels show the relative performances of combinations of distance metric (varying over columns of panels) and clustering methods (shown by the colours of the lines), as measured by the variation of information metric (y-axes), which is a measure obtained when comparing the inferred solution with the true solution (higher values show a larger departure from the correct solution). In each individual panel, the x-axis represents the number of NNI rearrangements between underlying clusters.*

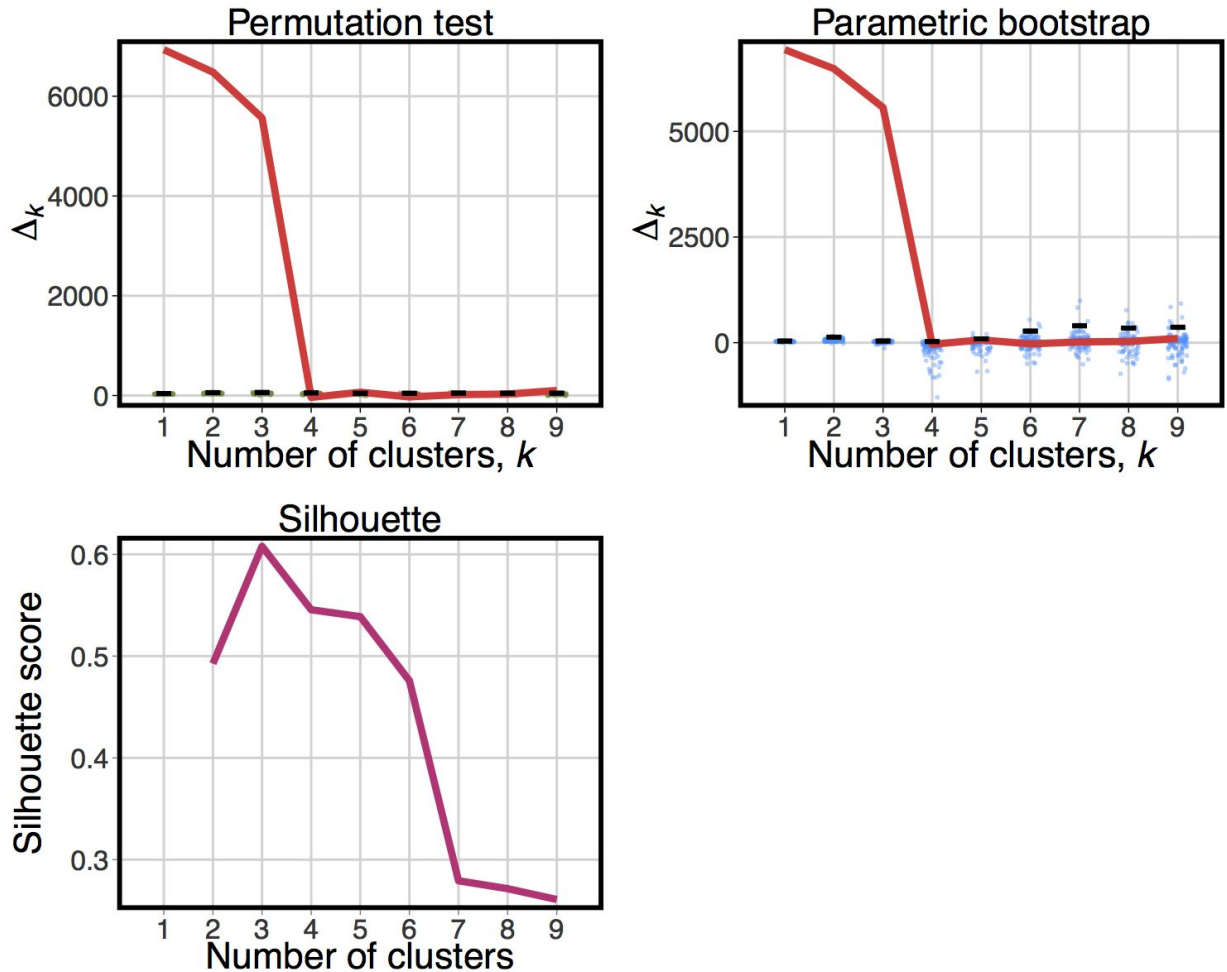

*Supplementary Figure 4: Comparison of the criteria used to determine the number of clusters on a single problem instance with true number of clusters equal to four. The simulation scenario is the same as in Figure 3, however this figure shows an instance for which the criteria do not all agree on the number of clusters. (A) Permutation test: the improvement in likelihood for each additional cluster (red curve) is significantly greater than that observed for permuted data sets (green dots show the distribution of values over 100 permutations) until the comparison between 4 and 5 clusters is reached, correctly implying that the use of 4 clusters is optimal. (B) Parametric bootstrap test: again, the improvement for each additional cluster (red curve) is significantly greater than that for data sets simulated for one fewer cluster (blue dots) until the true number of clusters (4) has been reached. (C) the silhouette score, a general-purpose stopping criterion, has its maximum at a value of 3. In this instance, the newly devised methods give the correct answer, while the general purpose silhouette criterion underestimates the number of clusters.*

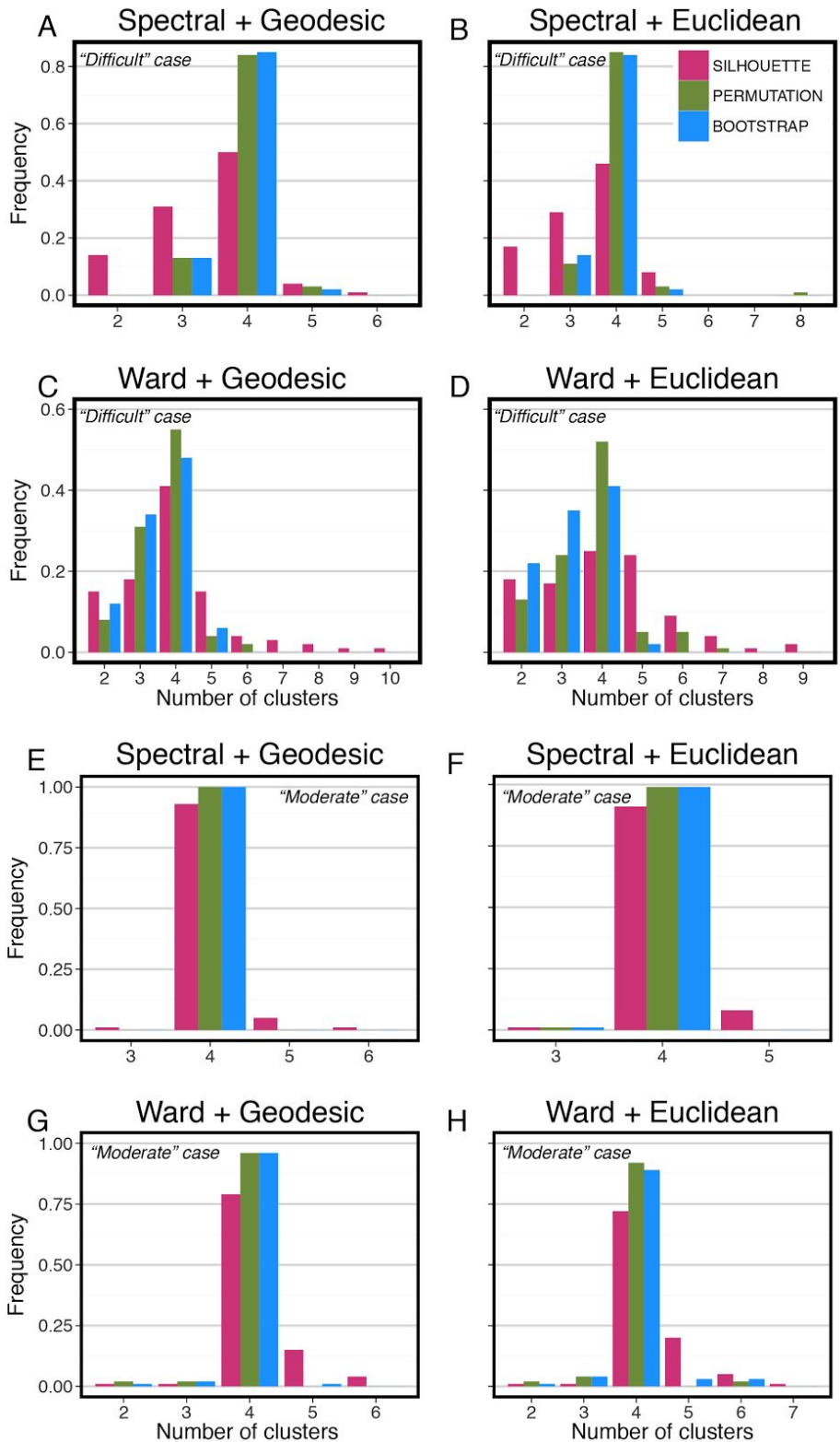

*Supplementary Figure 5: Distributions of the number of clusters found for 100 "difficult" problem instances, analysed under the four combinations of spectral / Ward's method clustering and Euclidean / geodesic distances. In every instance the true number of clusters is 4. In each separate case the special-purpose permutation and bootstrap methods outperform the general-purpose silhouette method at selecting the correct number of clusters. When wrong, all methods tend towards underestimation rather than overestimation. Spectral clustering outperforms Ward's method clustering, and geodesic distances slightly outperform Euclidean.*

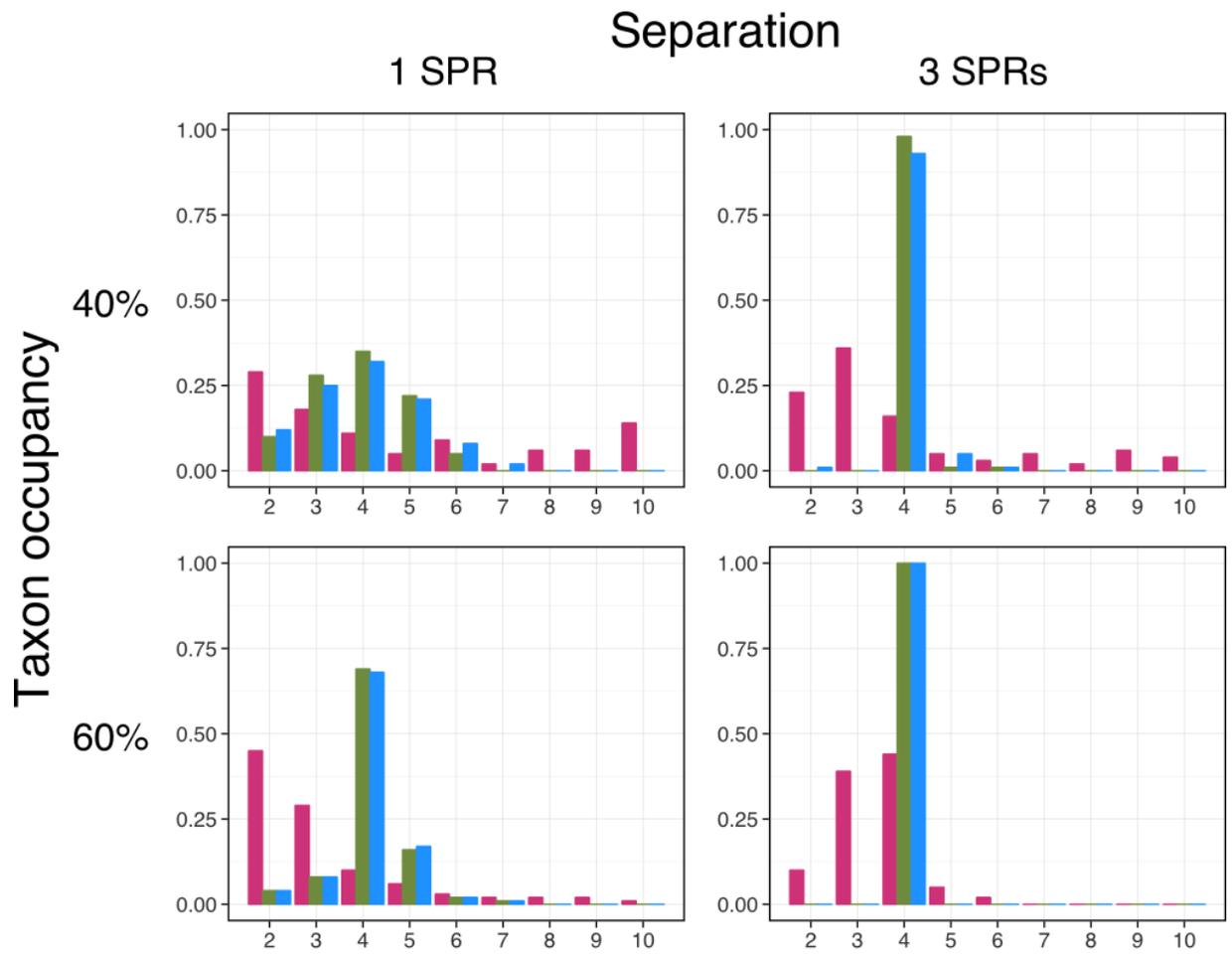

*Supplementary Figure 6: The performance of criteria for determining the number of clusters, with sparse data. Four examples are shown, for two levels of occupancy (40 and 60%; rows), and two levels of cluster separation (1 and 3 SPRs; columns). Occupancy is expressed as a percentage; 40%, for example, means that each taxon was included in any particular locus with probability 0.4.*

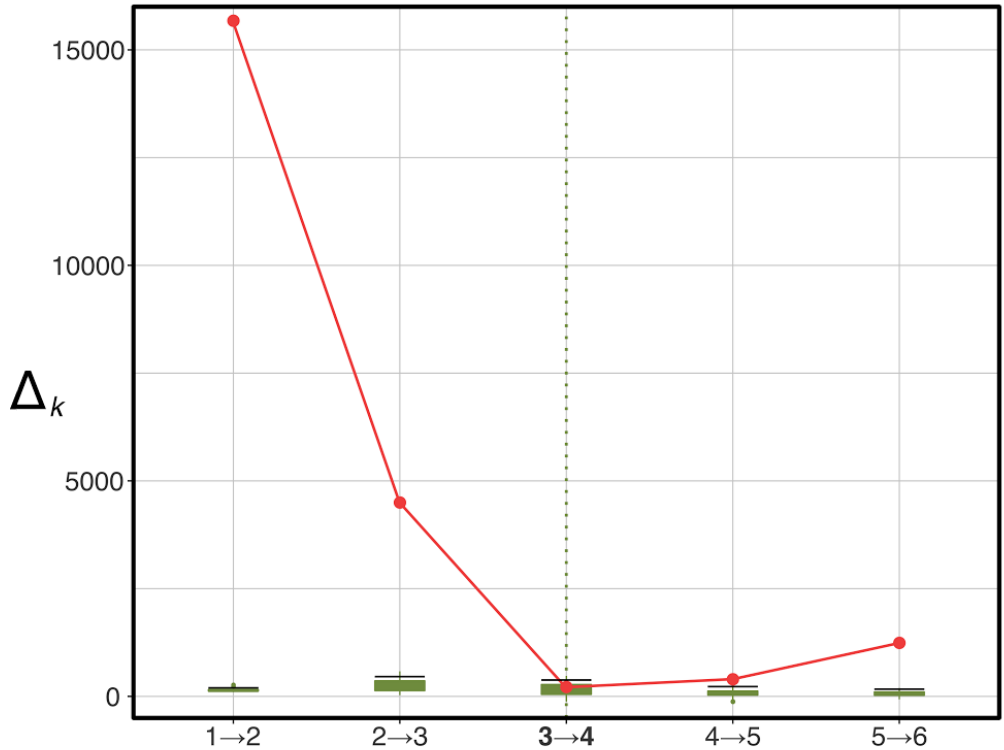

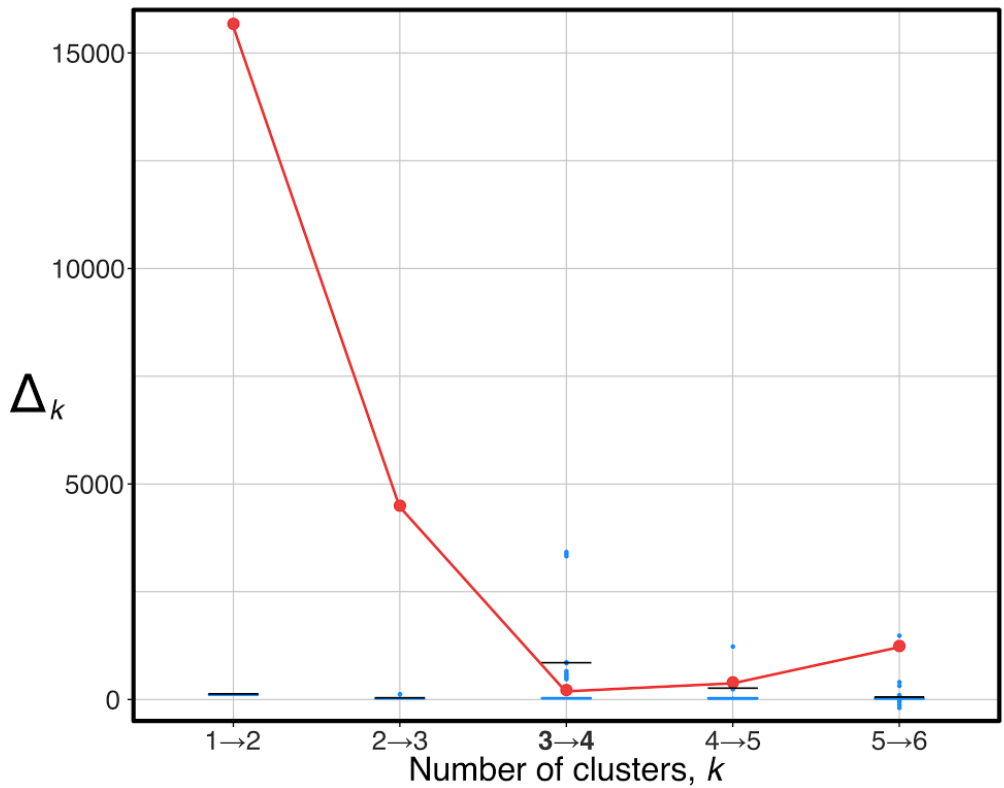

*Supplementary Figure 7: Stopping criteria applied to yeast dataset. The top-most panel shows the result of applying the permutation-based (non-parametric) variant of the stopping criterion to the yeast dataset. The lower panel shows the result of applying the parametric bootstrap variant of the stopping criterion. Both variants suggest that the data should be partitioned into three clusters.*

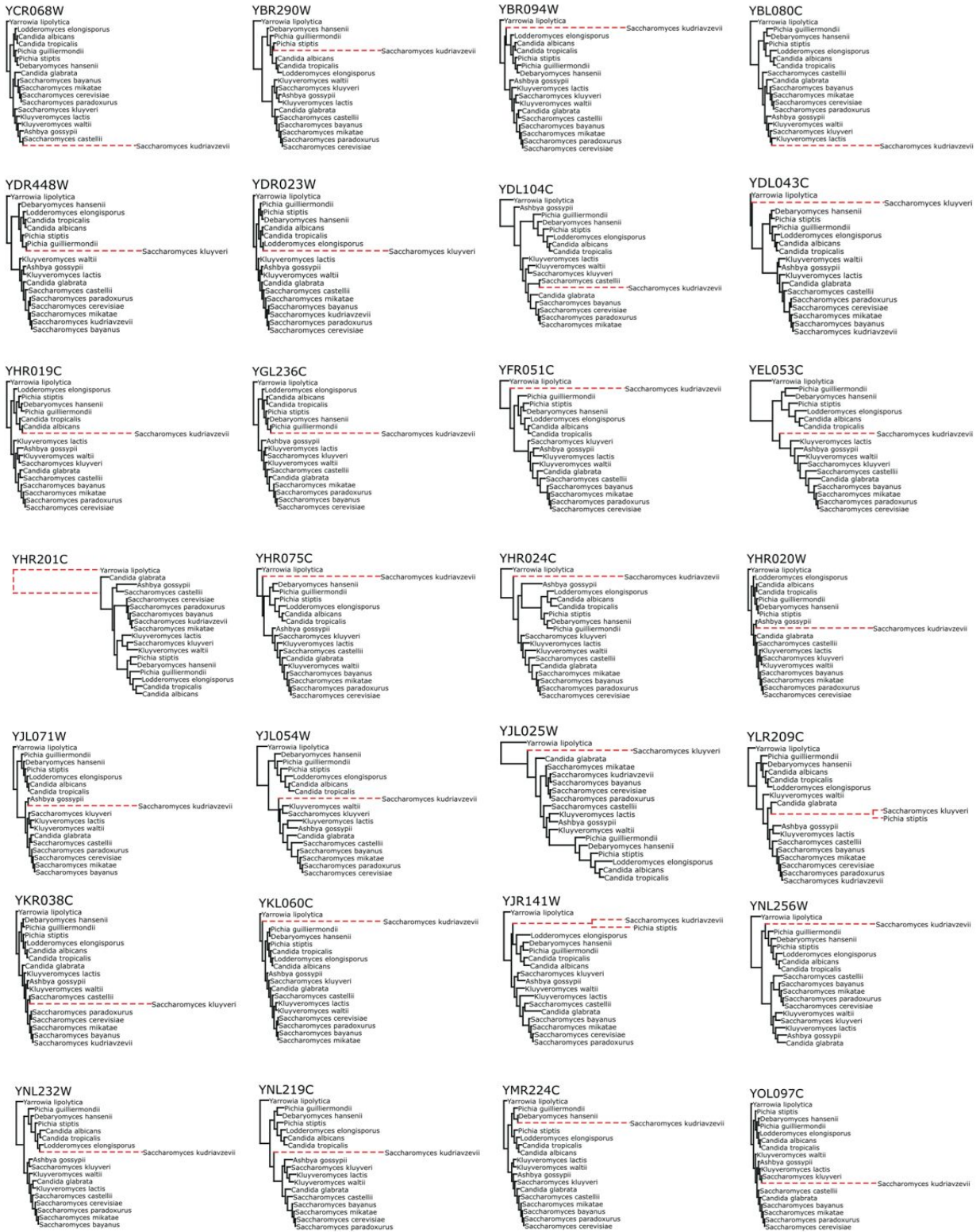

(*continued*...)

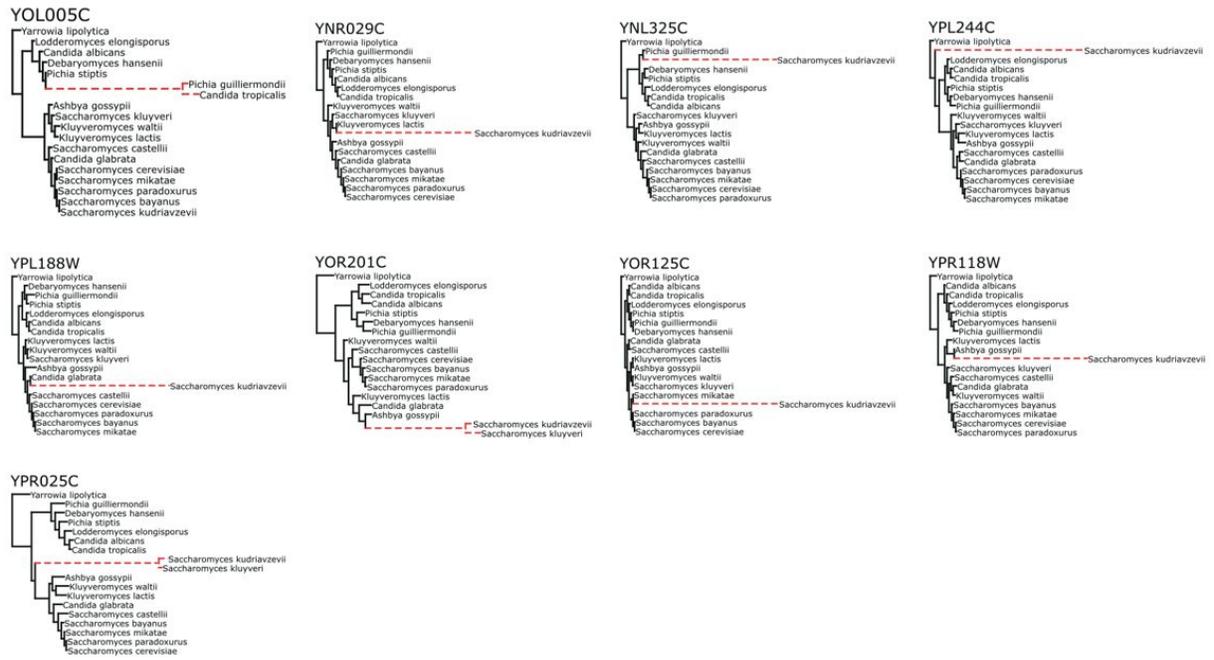

*Supplementary Figure 8: Phylogenetic trees for the 37 yeast loci discovered to have erroneous orthology, with the non-orthologous sequences highlighted as red dashed lines.*

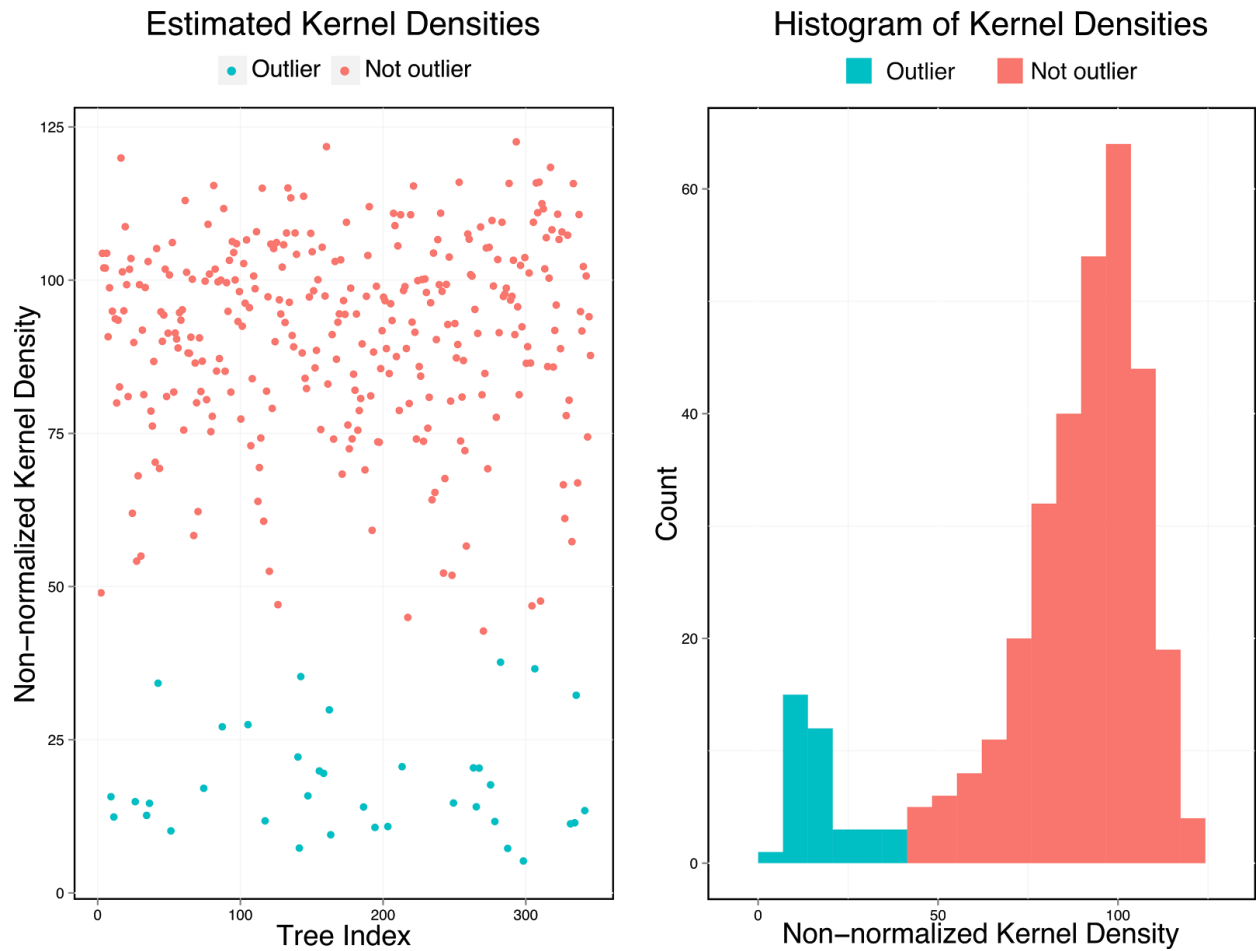

*Supplementary Figure 9: Application of kdetrees to the yeast dataset. The scatterplot in the left panel shows the kernel density estimate for each tree. The order of the trees along the x-axis is arbitrary with respect to cluster membership, rather being derived from the alphabetical ordering of the names of the loci. The right panel shows a histogram of the kernel density scores. In both panels, the outliers are coloured blue.*

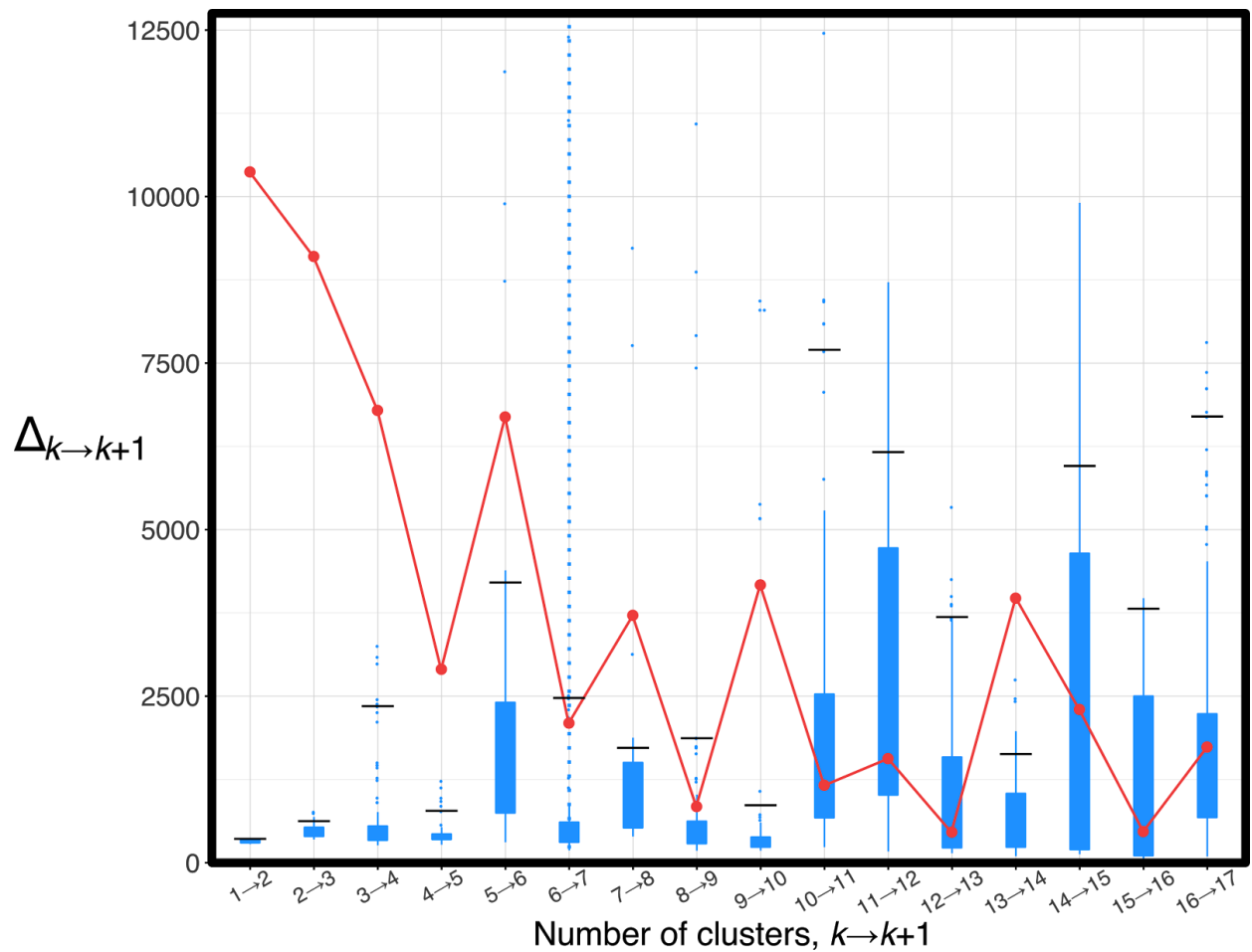

*Supplementary Figure 10: Likelihood improvement gained when partitioning the* Chiastocheta *data into increasing numbers of clusters (red points), using the parametric bootstrap criterion. The number of clusters selected by the stopping criterion is indicated by the vertical dashed line. Compared with the permutation test (Figure 8), the parametric bootstrap procedure yields a much larger variance in the likelihood improvement. Examination of the data reveals that this is due to the shallowness of some cluster trees, which makes it challenging for treeCl to identify the optimal clusters under both the null and alternative hypotheses. However, the conclusion (at least 4 clusters) is consistent with that of the permutation test.*

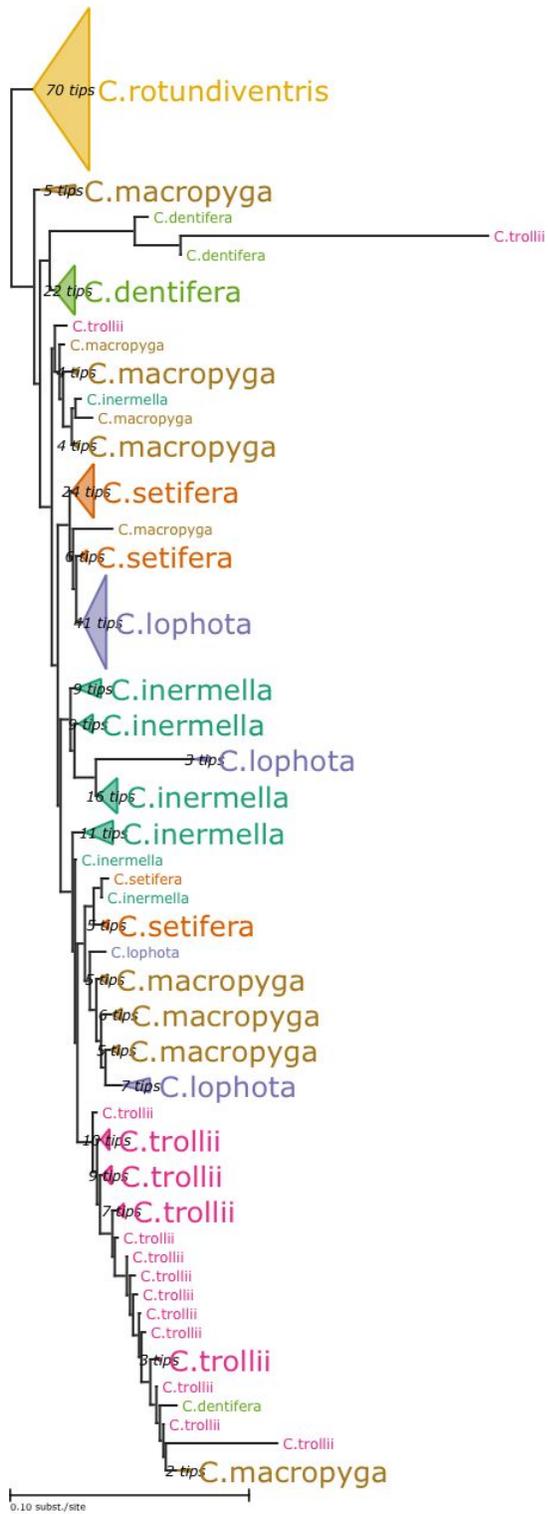

*Supplementary Figure 11: A tree obtained from one of the clusters when partitioning* Chiastocheta *loci into 5 clusters. The species are no longer monophyletic.*

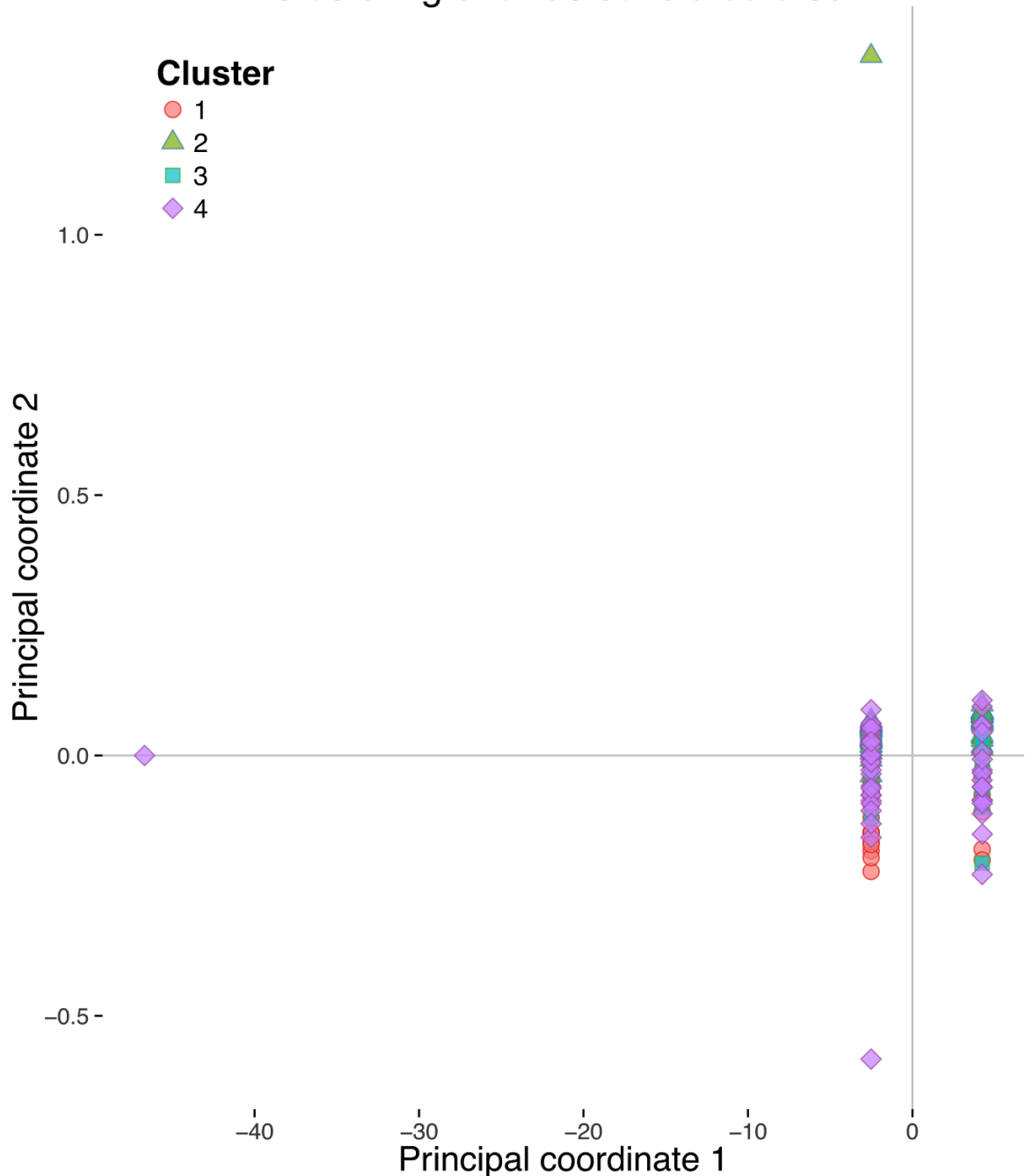

*Supplementary Figure 12: Embedding of chiastocheta trees using classical multidimensional scaling. With the exception of three outliers, the trees form two groups that are clearly separated by the first principal coordinate. However, this separation is not indicative of the cluster structure detected by treeCl using spectral clustering. Classical multidimensional scaling can be distorted when the input distances are not Euclidean (Torgerson 1952; Gower and Legendre 1986). In this case, non-Euclidean distances may arise from missing species' sequences for some loci potentially causing violations of the triangle equality amongst the inter-locus tree distances.*

| Orthologous Group | Misannotated Species | Best hit in S. cerevisiae |
|---|---|---|
| YBL080C | *Saccharomyces kudriavzevii* | YMR219W |
| YBR094W | *Saccharomyces kudriavzevii* | YLR357W |
| YBR290W | *Saccharomyces kudriavzevii* | YLR114C |
| YCR068W | *Saccharomyces kudriavzevii* | YJR107W |
| YDL043C | *Saccharomyces kluyveri* | YDL051W |
| YDL104C | *Saccharomyces kudriavzevii* | YKR038C |
| YDR023W | *Saccharomyces kluyveri* | YHR011W |
| YDR448W | *Saccharomyces kluyveri* | YFR037C |
| YEL053C | *Saccharomyces kudriavzevii* | YOL080C |
| YFR051C | *Saccharomyces kudriavzevii* | YPL259C |
| YGL236C | *Saccharomyces kudriavzevii* | YBL098W |
| YHR019C | *Saccharomyces kudriavzevii* | YCR024C |
| YHR020W | *Saccharomyces kudriavzevii* | YER087W |
| YHR024C | *Saccharomyces kudriavzevii* | YLR163C |
| YHR075C | *Saccharomyces kudriavzevii* | YLR133W |
| YHR201C | *Yarrowia lipolytica* | YMR052C-A |
| YJL025W | *Saccharomyces kluyveri* | YDR285W |
| YJL054W | *Saccharomyces kudriavzevii* | YBL052C |
| YJL071W | *Saccharomyces kudriavzevii* | YPR185W |
| YJR141W | *Saccharomyces kudriavzevii* | YLR019W |
| | *Pichia stipitis* | YNL144C-like |
| YKL060C | *Saccharomyces kudriavzevii* | YER043C |
| YKR038C | *Saccharomyces kluyveri* | YDL104C |
| YLR209C | *Saccharomyces kluyveri* | YLR017W |

|  | *Pichia stipitis* | YLR017W |
| --- | --- | --- |
| YMR224C | *Saccharomyces kudriavzevii* | YAL035W |
| YNL219C | *Saccharomyces kudriavzevii* | YGL142C |
| YNL232W | *Saccharomyces kudriavzevii* | YBL052C |
| YNL256W | *Saccharomyces kudriavzevii* | YPL070W |
| YNL325C | *Saccharomyces kudriavzevii* | YNL106C |
| YNR029C | *Saccharomyces kudriavzevii* | YPL009C |
| YOL005C | *Candida tropicalis* | YNL113W |
|  | *Pichia guilliermondii* | YNL113W |
| YOL097C | *Saccharomyces kudriavzevii* | YGR185C |
| YOR125C | *Saccharomyces kudriavzevii* | YER086W |
| YOR201C | *Saccharomyces kudriavzevii* | YLR051C |
|  | *Saccharomyces kluyveri* | YLR051C |
| YPL188W | *Saccharomyces kudriavzevii* | YKR056W |
| YPL244C | *Saccharomyces kudriavzevii* | YEL004W |
| YPR025C | *Saccharomyces kudriavzevii* | YNL025C |
|  | *Saccharomyces kluyveri* | YNL025C |
| YPR118W | *Saccharomyces kudriavzevii* | YKR026C |

*Supplementary Table 1: Summary of erroneous orthology discovered in the yeast data set. The first column gives the orthologous group to which the sequences from the species in the second column were assigned. The third column gives the gene name of the best BLAST hit in* S. cerevisiae.